\documentclass[12pt,preprint]{aastex}

\begin{document}

\title{Bars in Disk-Dominated and Bulge-Dominated Galaxies at $z\sim0$:
New Insights from $\sim 3600$ SDSS Galaxies}

\author{Fabio D. Barazza\altaffilmark{1}, Shardha Jogee, and Irina Marinova}
\affil{Department of Astronomy, University of Texas at Austin, 1 University
Station C1400, Austin, TX 78712-0259, USA}
\altaffiltext{1}{current address: Laboratoire d'Astrophysique, \'Ecole
Polytechnique F\'ed\'erale de Lausanne (EPFL), Observatoire, CH-1290 Sauverny,
Switzerland}
\email{barazza@astro.as.utexas.edu, sj@astro.as.utexas.edu,
marinova@astro.as.utexas.edu}

\begin{abstract}
We present a study of large-scale bars in the local Universe, based on a large
sample of $\sim3692$  galaxies,  with $-18.5\leq M_g < -22.0$ mag and redshift
$0.01\leq z<0.03$, drawn from the Sloan Digitized Sky Survey. While most
studies of bars in the local Universe have been based on relatively small
samples that are dominated by bright galaxies of early to intermediate
Hubble types with prominent bulges, the present sample is $\sim$~10 times 
larger, covers a larger volume, and includes many galaxies that are 
disk-dominated and of late Hubble types. Both color cuts and S\'ersic
cuts yield a similar sample of $\sim2000$ disk galaxies. We
characterize bars and disks by ellipse-fitting  $r$-band images
and applying quantitative criteria. After excluding highly inclined
($>60^{\circ}$) systems, we find the following results. (1)~The optical
$r$-band fraction ($f_{\rm opt-r}$) of barred galaxies, when averaged over the
whole sample, is $\sim48\%-52\%$. The bars have diameters $d$ of 2 to 24 kpc,
with most ($\sim72\%$) having $d\sim$ 2 to 6 kpc. (2)~When galaxies are
separated according to half light radius ($r_{\rm e}$), or normalized
$r_{\rm e}$/$R_{\rm 24}$, which is a measure of the bulge-to-disk ($B/D$)
ratio, a remarkable result is seen: $f_{\rm opt-r}$ rises sharply,
from $\sim$~40\% in galaxies that have small $r_{\rm e}$/$R_{\rm 24}$ and
visually appear to host prominent bulges, 
to $\sim$~70\% for galaxies that have large $r_{\rm e}$/$R_{\rm 24}$ and
appear disk-dominated. Visual classification, performed for $\sim900$
galaxies, confirms our result that disk-dominated galaxies
with no bulge or a very low $B/D$ display a significantly higher optical bar
fraction ($>70$\% vs 40\%) than galaxies with prominent bulges. It also shows
that barred galaxies host a larger fraction (31\% vs 5\%) of quasi-bulgeless
disk-dominated galaxies than do unbarred galaxies. (3)~$f_{\rm opt-r}$
rises for galaxies with bluer colors (by $\sim30\%$) and lower masses
(by $\sim15\%-20\%$). (4) The significant rise in the
optical bar fraction toward late-type galaxies is discussed in terms
of their higher gas mass fraction, higher dark matter fraction, and
lower bulge-to-disk ratio. (5) While hierarchical
$\Lambda$CDM models of galaxy evolution models fail to produce
galaxies without classical bulges, our study finds that $\sim20\%$
of disk galaxies  appear to be ``quasi-bulgeless''. (6)~Our
study of bars at $z\sim0$ in the optical $r$-band provides the
$z\sim0$ comparison point for $HST$ ACS surveys (e.g., GEMS, GOODS,
COSMOS) that measure the rest-frame optical bar fraction in
bright galaxies out to $z\sim1$. After applying the same cutoffs in
magnitude, bar ellipticity ($e_{\rm bar}\ge0.4$), and bar size
($a_{\rm bar}\ge1.5$ kpc), which are applied in $z\sim0.2-1.0$
studies in order to trace strong bars with adequate spatial resolution 
in bright disks, we obtain an optical $r$-band bar fraction of $34\%$.
This is comparable to the value reported at $z\sim0.2-1.0$, implying
that the optical fraction of strong bars does not suffer a dramatic
order of magnitude decline in bright galaxies out to $z\sim1$.
\end{abstract}


\section{Introduction}\label{intro}
The majority ($\sim$~60\%) of bright disk galaxies are barred, when observed in
the near-infrared \citep[][hereafter MJ07]{kna99,esk00,lau04,men07,mar07} and a significant
fraction of these ($\sim$~45\%) also appear barred in the optical \citep[][MJ07]{esk00}. 
Earlier studies suggested a striking or order of magnitude decline in
the optical fraction of bars out to $z\sim1$ \citep{abr99,vdb00},
but subsequent studies have ruled out an order of magnitude decline
and find that the optical fraction of strong bars remains fairly
constant or show a moderate decline of a factor of $\sim2$
\citep[][Sheth et al. 2003,2004,2007; see $\S$ \ref{discu}]{jog04,elm04a,zhe05}.

Bars are believed to be very important with regard to the dynamical and secular
evolution of disk galaxies, particularly in redistributing the angular momentum
of the baryonic and dark matter components of disk galaxies
\citep{com81,wei85,com90,deb00}. The interaction between the bar and the disk
material can lead to the inflow of gas from the outer disk to the central
parts, which can trigger starbursts
\citep{elm94,kna95,reg04,jog05,sht05} and might
contribute to the formation of disky bulges
\citep[or 'pseudobulges',][]{kor93,kor04,ath05a,jog05,deb06}.
Additional evidence for secular evolution is provided by box- or
peanut-shaped bulges in inclined galaxies. These features are
commonly attributed to the orbital structure, resonances, and vertical
instabilities in a barred potential \citep{com90,kui95,bur99,bur05}.

From a theoretical perspective, it is possible to model some aspects of the
evolution of disks and bars, and their interactions (e.g., the corresponding
simulations are able to reproduce certain broad features of barred disks).
However, it remains unclear why a specific galaxy has a bar, but a seemingly
similar galaxy is unbarred; or why some barred galaxies have a classical bulge,
whereas others harbor a disky bulge, etc. This might indicate that specific
properties of the disks or the particular processes involved in their formation
have a strong impact on their ability to form a bar. In order to investigate,
how disk and bar formation are related, it is not only important to determine
the fraction of disk galaxies that are barred, but also to relate bar and disk
properties. There are different methods to find and characterize bars. The
Third Reference Catalog of Bright Galaxies \citep[][hereafter RC3]{dev91} uses 
three bar strength families (SA, SAB, and SB) to characterize bars
based on a visual inspection of blue light images. Using this classification
\cite{ode96} showed that the optical fraction of strong bars in disk
galaxies rises from Sc galaxies towards later-types. More
quantitative measures, such as the gravitational torque method
\citep{blo02,lau02,but05}, or Fourier dissection \citep{but06,lau06}, were also
used, not only to find bars, but to quantitatively determine bar strengths and
bar lengths. Similarly, the method of fitting ellipses to galaxy isophotes
provides a tool to characterize the length and shape of bars
\citep{fri96,jog99,kna00,sht00,lai02,why02,jog02a,jog02b,sht03,elm04a,ree07,mar07,men07,sht07}.

These efforts were able to shed light on the fraction, shapes, and structures 
of bars  in local disk galaxies of early to intermediate Hubble types. 
First attempts were made to relate the presence of a bar or its
structural properties to other galaxy characteristics. However, there were
three important limitations. Firstly, samples used in earlier studies were
small ($\sim100$ to 200 objects) and mostly composed of bright
galaxies of early to intermediate Hubble types (Sa to Sc), with fairly 
prominent bulges. One could barely get decent number statistics for
bars in early-type disk galaxies, while the bins of disk-dominated
late Hubble types were dominated by Poisson noise
(e.g., see Figure 16 in MJ07). Secondly, with such small samples, it
was difficult to bin galaxies in terms of the galaxy host
properties. Thirdly, earlier samples were drawn from a very small
volume, and could be highly impacted by cosmic variance.

In the present study, we use a sample of $\sim2000$ galaxies, at $z=0.01-0.03$
with $M_r \sim -18.5$ to $-22.0$ mag. The first advantage of this study is that
it provides a factor of 10 improvement in number statistics and reduces the
effect of cosmic variance by selecting galaxies drawn from a larger volume. 
Secondly, with $\sim2000$ galaxies, we can for the first time have  $100-200$ 
galaxies per bin, while binning galaxies in terms of host galaxy parameters, 
such as  luminosity, measures of bulge-to-disk ($B/D$) ratios, size, colors,
surface brightness, etc. This allows us to  conduct a comprehensive study 
of barred and unbarred galaxies as a
function of host galaxy properties. Thirdly, our sample has a large number of
galaxies, which are relatively faint ($M_g>-19.5$ mag) or/and appear
disk-dominated, characteristic of late Hubble types. This allows us to
shed light on what happens to bars at the fainter end of the
luminosity function and in the regime of disk-dominated galaxies.

A fourth goal of our study is to provide a reference baseline for bars at
$z\sim0$ in the {\it rest-frame optical} for intermediate redshift $HST$
surveys using the Advanced Camera for Surveys (ACS), such as the Tadpole field
\citep{tra03}, the Galaxy Evolution from Morphologies and SEDs
\citep[GEMS,][]{rix04}, the Great Observatories Origins Deep Survey
\citep[GOODS,][]{gia04}, and COSMOS \citep{sco06}, which trace bars in the
rest-frame optical band at $z\sim0.2-1.0$ (look-back times of 3--8 Gyr). We use
SDSS to provide the reference point at $z=0$ in the $r$-band, complementing the
one in the $B$-band of MJ07. Our $B$- and $r$-band results can be directly
compared to $HST$ ACS optical studies of bars in bright disks at $z\sim0.2-1.0$
\citep{elm04a,jog04}. The validity of this comparison is reinforced by the fact
that we use the same procedure of ellipse fits ($\S$ \ref{meth}) that were used
by these studies. We also note that the reference $z=0$ point for bars in the
near-infrared band \citep{men07} is not appropriate for comparison with the
above $HST$ ACS surveys, which trace the  rest-frame optical rather than  the 
rest-frame near-infrared. 

The outline of the paper is as follows: In $\S$ \ref{samsel} we present our
sample selection. The method used to find and characterize bars is explained in
$\S$ \ref{meth}. In $\S$ \ref{resol} we discuss the detection limits. Our
results and more detailed assessments of specific findings are presented in
$\S$ \ref{resul}. We discuss our results in $\S$ \ref{discu} and summarize our
conclusions in $\S$ \ref{sum}. Throughout the paper we assume a flat cosmology
with $\Omega_M=1-\Omega_{\Lambda}=0.3$ and $H_0=70$ km~s$^{-1}$ Mpc$^{-1}$.

\section{Sample selection}\label{samsel}
Our sample of local disk galaxies is drawn from the low-redshift catalog of the
New York University Value-Added Catalog \citep[NYU-VAGC,][]{bla05}. The
NYU-VAGC is based on the second data release of SDSS, which is acquiring {\it
ugriz} CCD imaging of $10^4$ deg$^2$ of the northern Galactic sky and selecting
$10^6$ targets for spectroscopy, most of them galaxies with $r<17.77$ mag
\citep{aba04}. The low-redshift catalog consists of 28089 galaxies at distances
of 10--200 Mpc ($0.0033<z<0.05$), which have been determined by correcting for
peculiar velocities. For each galaxy, background subtracted and deblended
images in $ugriz$ as well as individual PSF-frames are available.

\begin{figure}
\epsscale{1}
\plotone{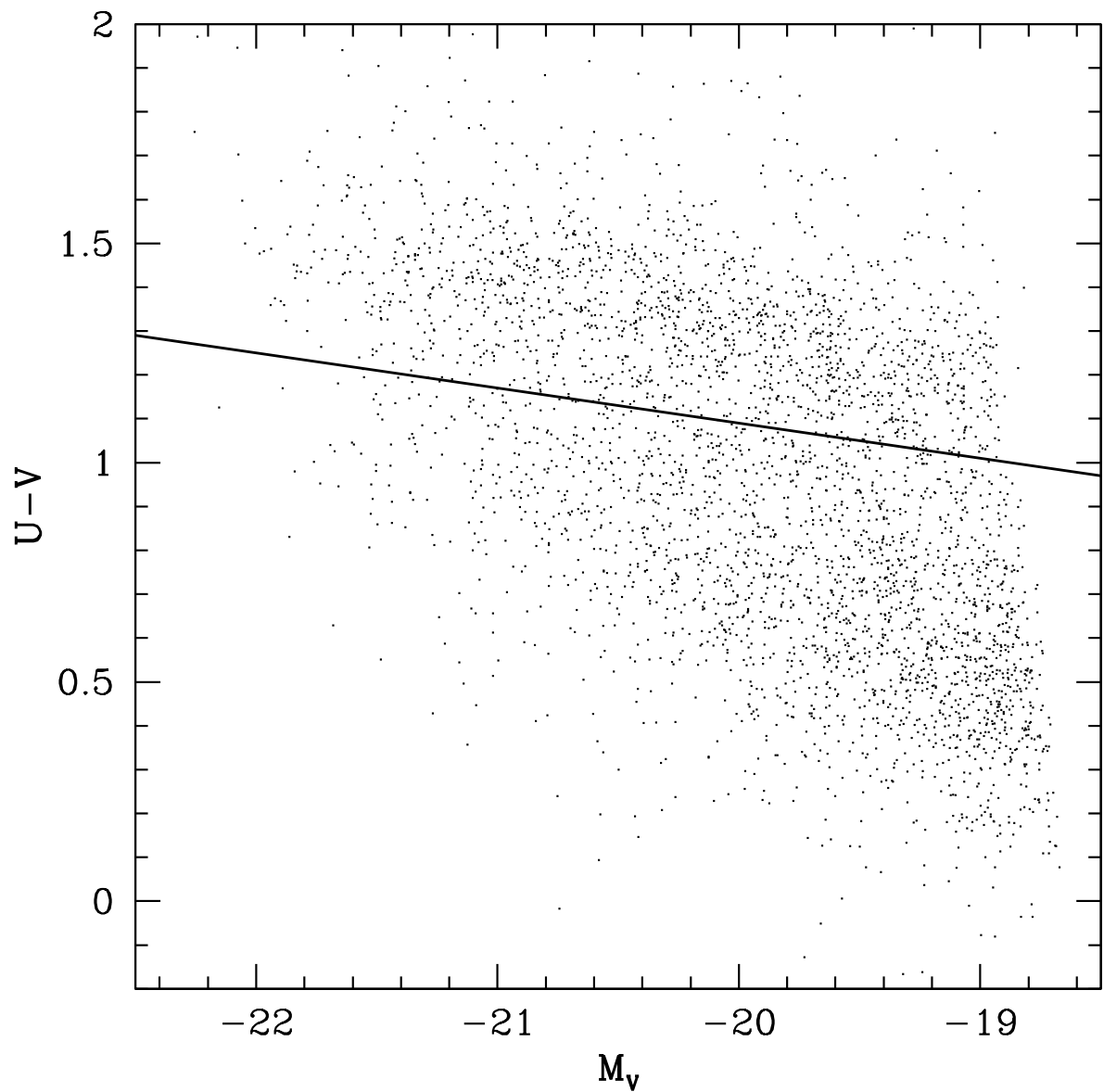}
\caption{The color-magnitude diagram of our initial sample of 3692 galaxies.
The solid line corresponds to $U-V=1.15-0.31z-0.08(M_V-5\log(h)+20)$
\citep{bel04a}. The blue galaxies lying below this line (1961 objects) are
included in our sample of local disk galaxies.\label{cmd}}
\end{figure}

We selected galaxies in the redshift range $0.01<z<0.03$, having $M_g\leq-18.5$
mag. The typical seeing ($1\farcs4$) corresponds to 290 to 840 pc over
$0.01\leq z<0.03$. This is adequate for  resolving large-scale bars, which are
the focus of this study.  Large-scale bars have diameters $\ge$ 2 kpc and their
lengths encompass at least 2.5 independent PSFs, allowing them to be
resolved and fitted. We note that bars with diameters below 2 kpc or semi
major axis $a_{bar}<1$ kpc are typically considered as nuclear rather than
large-scale bars \citep{lai02}. Nuclear bars are not the focus of this study
and are excluded from our results. Hence, it is not a source of concern, if
some of them are unresolved by the data. The selected sample is complete down
to $M_g\leq-18.5$ mag in the chosen redshift range and consists of 3692
objects. The magnitude distribution covers $M_r = -18.5$ to $-22.0$ mag (Figure
\ref{cmd}).

We use the GALFIT \citep{pen02} software to perform single component S\'ersic
fits of the form
\begin{displaymath}
\Sigma(r)=\Sigma_e\exp\left[-\kappa\left(\left(\frac{r}{r_{\rm e}}\right)^{1/n}-1\right)\right]
\end{displaymath}
to the 2D-images, which provide S\'ersic indices ($n$) and half-light radii
($r_{\rm e}$) for the galaxies. $\Sigma_e$ is the pixel surface brightness at
$r_{\rm e}$ and $\kappa$ is a dependent variable and coupled to $n$. 

\begin{figure}
\epsscale{1}
\plotone{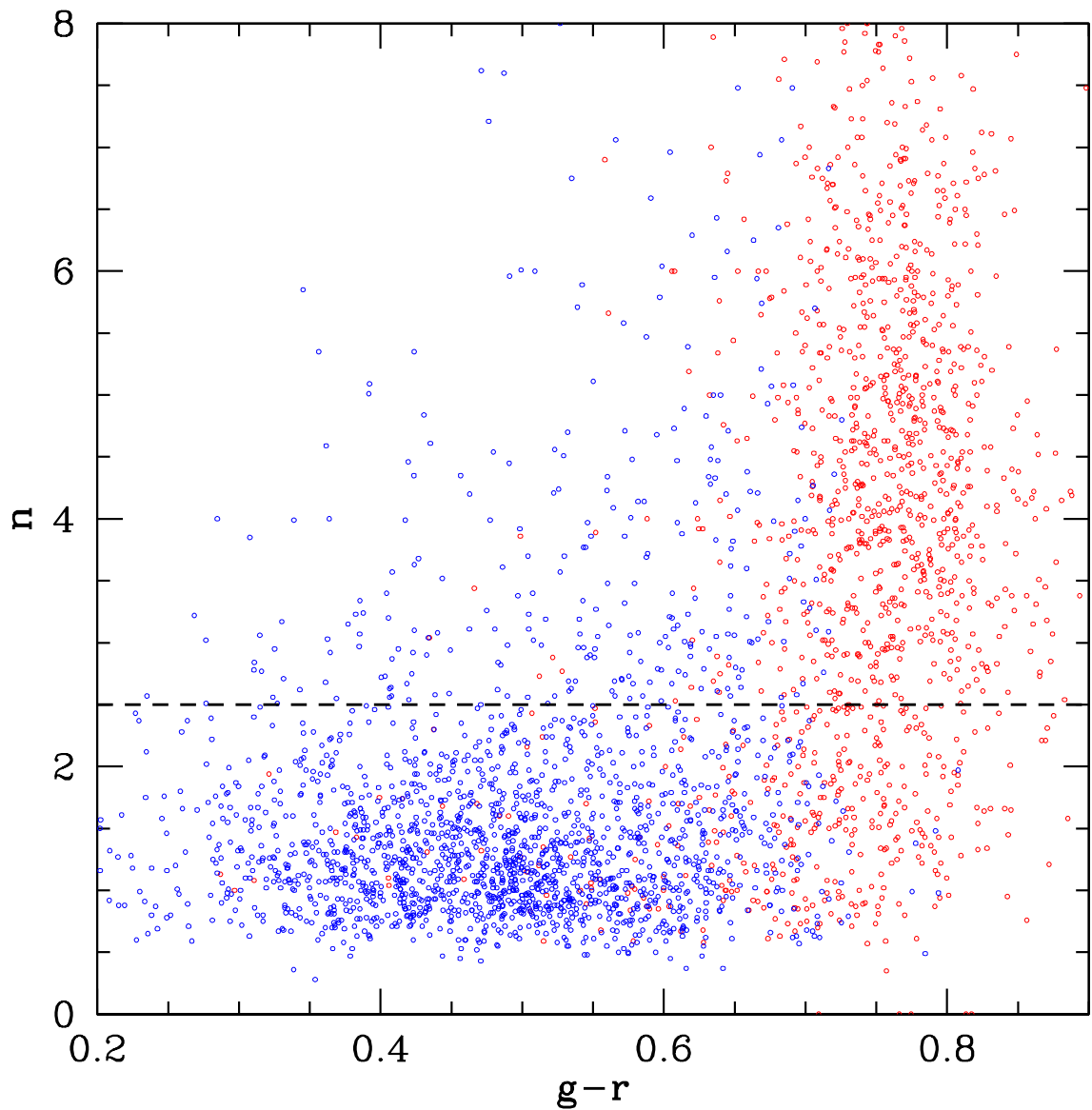}
\caption{Comparison of the subsamples based on S\'ersic and color cuts. The
$g-r$ color is plotted versus the S\'ersic index $n$. In the subsample based on
a color cut, the galaxies represented by blue dots are selected, while the red
dots are  excluded. The subsample based on a S\'ersic cut $n<2.5$ lie below the
dashed line. Blue galaxies below the line belong to both subsamples. Notice the
strong overlap between the two subsamples.
\label{cuts}}
\end{figure}

\begin{figure}
\epsscale{1}
\plotone{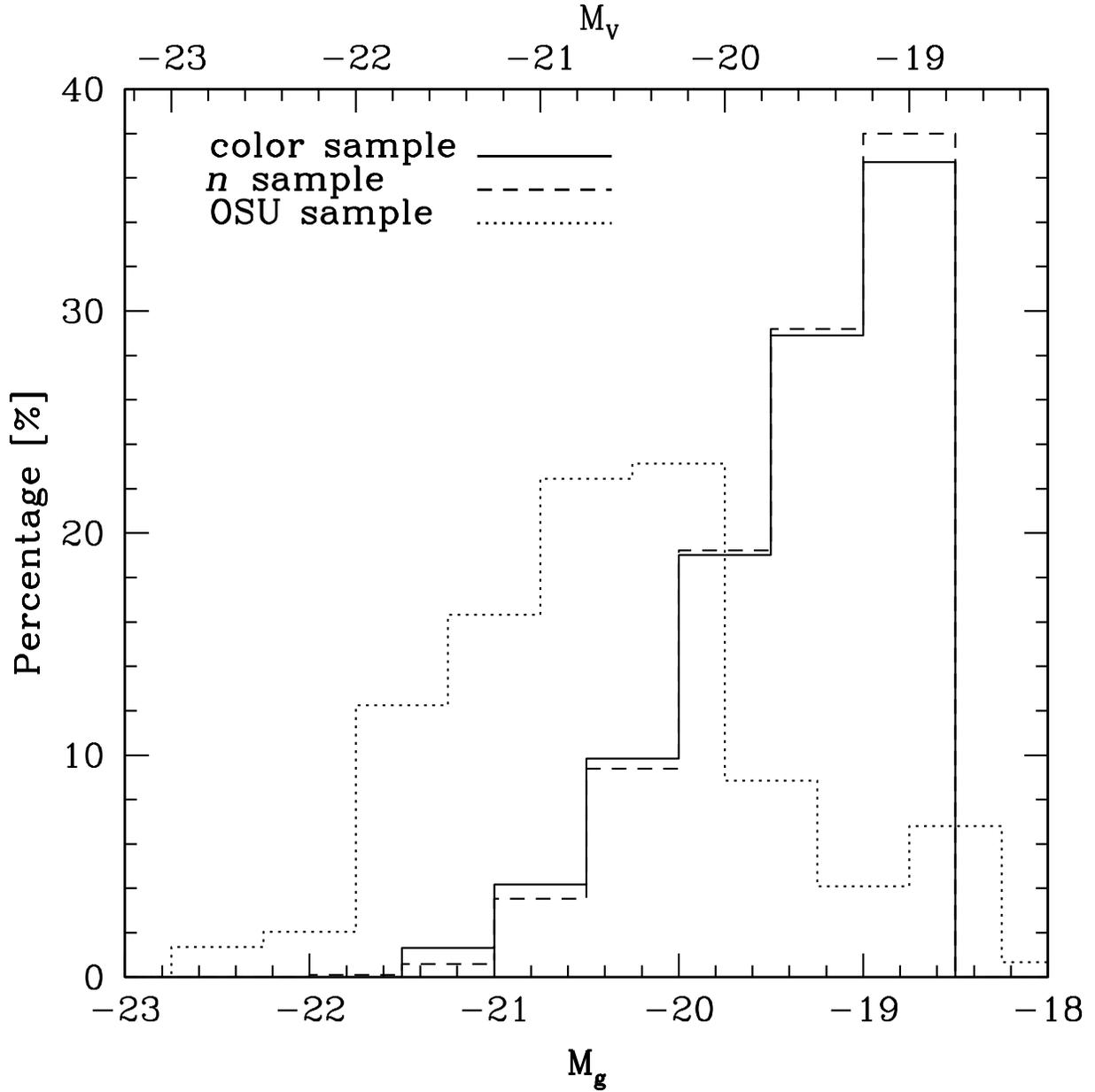}
\caption{The distribution of absolute magnitudes is shown for our subsamples of
disk galaxies based on S\'ersic and color cuts, and, as comparison, for the OSUBSGS
disk sample. For the latter we show $M_V$, which is given on the upper axis.
The strong dominance of fainter galaxies in the SDSS samples is
evident.\label{magd}}
\end{figure}

The optical bar fraction is defined as the sample of disk galaxies that host
large-scale bars, hence we must first define a sample of disk systems. There
are two common methods to separate disks and elliptical galaxies. For giant
galaxies, the first method  is to use the S\'ersic index from single component
fits and define giant disk galaxies to have $n<2.5$
\citep[e.g.][]{jog04,bel04b,bar05}. The second method is to apply a color cut
defined in color-magnitude space \citep[e.g.][]{bel04b,mci05,wol05}, assuming
that disk galaxies are predominantly bluer, star-forming systems. Both methods
have limitations. The color cut may miss out disk systems that are red due to
the presence of a dusty starburst. The S\'ersic cut may be contaminated by some
bright and rather blue dwarf ellipticals and miss out certain galaxies with
point-like AGN sources. We applied both methods to our sample as illustrated in
Figure \ref{cmd} and \ref{cuts}. The corresponding disk samples have an overlap
of $\sim85\%$ (Figures \ref{cuts} and \ref{magd}). In Figure \ref{cuts} the
color-selected and S\'ersic-selected samples are plotted in the $g-r$ versus
$n$ plane, with the color-selected galaxies in blue, and the good overlap is
evident. Figure \ref{magd} shows the similar magnitude distributions of the
color-selected and S\'ersic-selected disk samples from SDSS. Figure \ref{magd}
also illustrates a crucial property of the SDSS disk sample. The sample covers
a magnitude range of $\sim3$ mag, but is clearly dominated by fainter galaxies
($M_r = -18.5$ to $-20.0$ mag). Furthermore, visual inspection shows that a
large fraction of these galaxies seems to be disk-dominated, with little or no
bulge visible.  We do not have Hubble types for these SDSS galaxies, but these
characteristics are typical of late type galaxies (Sd, Sm). This is a crucial
point and has to be kept in mind for the discussion of the results presented
below. As discussed in $\S$ \ref{intro}, such galaxies are underrepresented in
most studies of bars in local disk galaxies carried out to date. As an
illustration, the magnitude distribution in $M_V$ of spirals in the 
OSU Bright Spiral  Galaxy Survey \citep[OSUBSGS,][]{esk02} is overplotted in
Figure \ref{magd}. This sample is often used as a reference sample for bars at
$z=0$ \citep[e.g.][MJ07]{esk00,blo02,why02,but05}. 

We finally opted for the color cut, since the contamination of $n<2.5$ objects
in the sample is slightly smaller than the fraction of red sequence galaxies in
the sample defined by $n<2.5$. We applied the color cut
$U-V<1.15-0.31z-0.08(M_V-5\log(h)+20)$ \citep{bel04a}, where $h=H_0/100$ and
$h=0.7$ is used in this paper. This cut is parallel to the red sequence in the
color-magnitude diagram and stems from an empirical fit to the evolution of the
color-magnitude relation for galaxy clusters at different redshifts. It
therefore corresponds to a definition of the red sequence based on a number of
nearby and distant clusters, but shifted by 0.25 mag to the blue. The
color-magnitude diagram for our sample is shown in Figure \ref{cmd}. The solid
line indicates the color cut. The resulting sample of disk galaxies contains
1961 objects.

The sample in this paper only includes galaxies in the redshift range $0.01\leq
z<0.03$. For completeness, we mention that the analysis outlined in $\S$
\ref{resul} has also been performed on the 1890 galaxies in the redshift range 
$z=0.03$ to 0.04, and yields essentially the same results. We did not include
these galaxies in the sample of this paper because the seeing at $z\sim$~0.04
($1\farcs4$ or 1.1 kpc) is in the limiting range where we can still resolve
large-scale bars of diameter $\ge$~2 kpc, but the fit may not be as robust. 

\section{Characterization of bars and disks}\label{meth}
The method used to find and characterize bars in disk galaxies is based on
fitting ellipses to the isophotes on the $r$-band images of our sample
galaxies, along with a set of quantitative criteria outlined below. We 
opted for the $r$-band, because it provides deeper images than
observations in the other SDSS filters. Many observational studies
have used and refined the method of fitting ellipses to characterize
bars \citep{fri96,jog99,kna00,sht00,lai02,why02,jog02a,jog02b,
sht03,elm04a,ree07,mar07,men07,sht07}. There is also a
strong body of theoretical evidence \citep{ath92a,she04} supporting this
approach, as outlined in MJ07. In particular, \cite{ath92a} studied orbits in
analytic potentials of barred galaxies and showed that generalized ellipses are
a good representation of the main bar-supporting stellar orbits. The departure
of ellipses fitted by the IRAF task 'ellipse' from these generalized ellipses
is characterized by the value of the harmonic amplitudes A3, B3, A4, and B4,
and we find that they are small (typically $<10\%$).

\begin{figure}
\epsscale{1}
\plotone{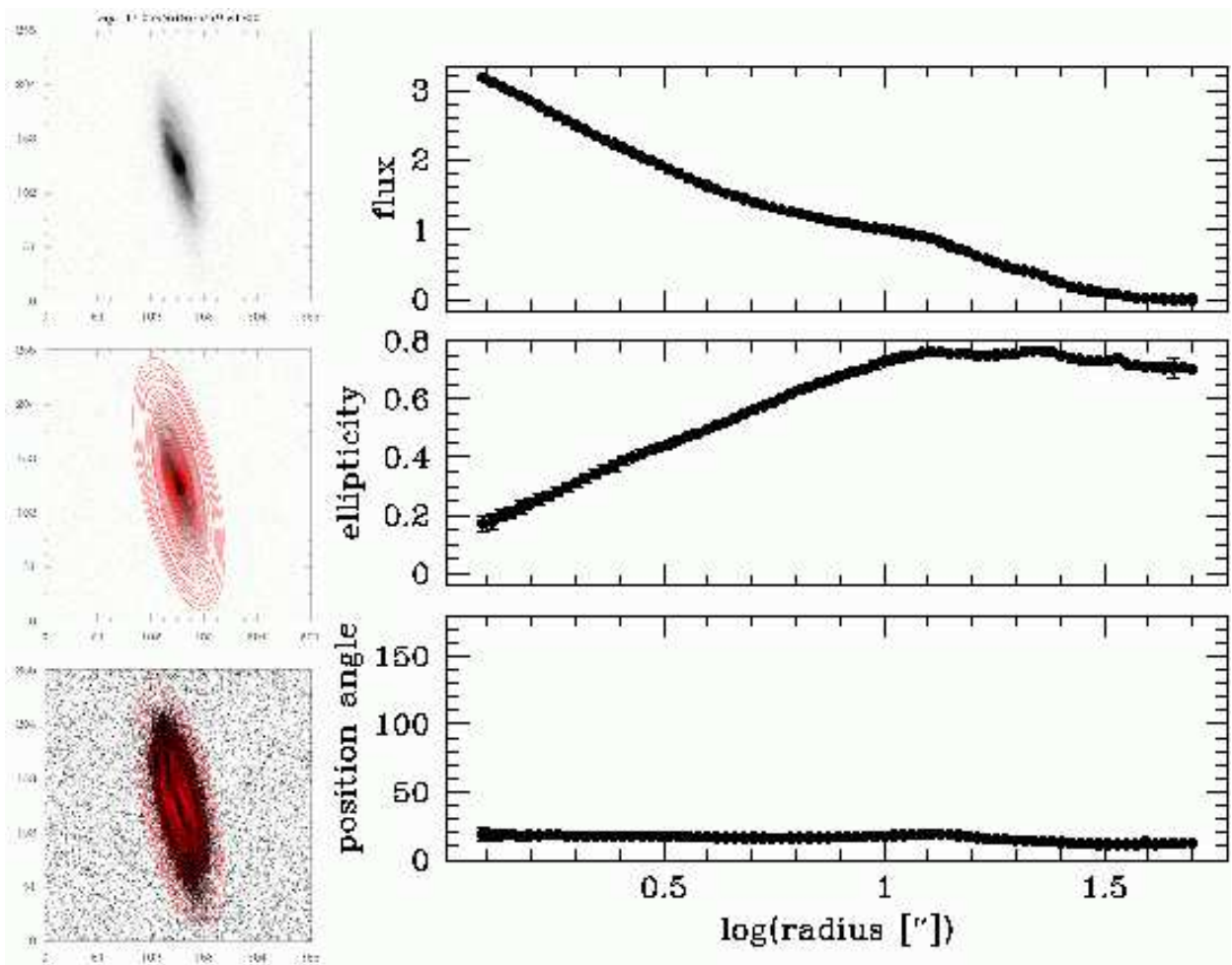}
\caption{Above is an example of an inclined ($i>60^{\circ}$) galaxy, which is
identified from the overlays and radial profiles generated by the ellipse fits.
Left panel: The top image shows only the galaxy, while the middle and bottom
images show the ellipses overlaid on the galaxy, with greyscale stretches
chosen to emphasize the inner (middle image) and outer (bottom image) regions
of the galaxy. The images are roughly $100\arcsec$ on a side. Right panel: The
radial profiles of surface brightness (top), ellipticity $e$ (middle), and PA
(bottom) are shown. In the outer parts of the galaxy, the PA is flat and the
ellipticity is fairly constant at $e >0.5$, indicating that the galaxy has a
large inclination $i>60^{\circ}$.
\label{pincl}}
\end{figure}

\begin{figure}
%\epsscale{2.3}
\plottwo{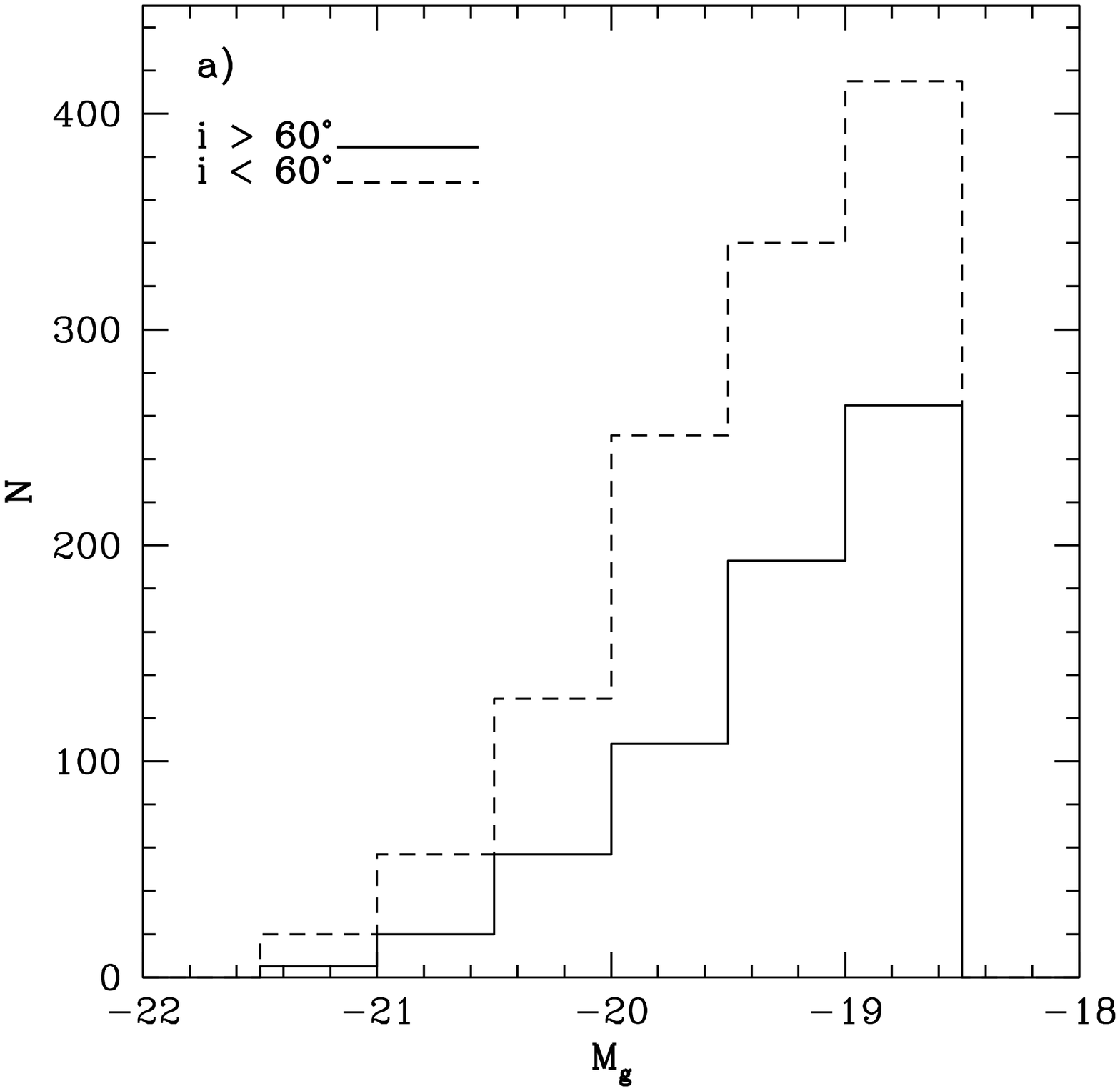}{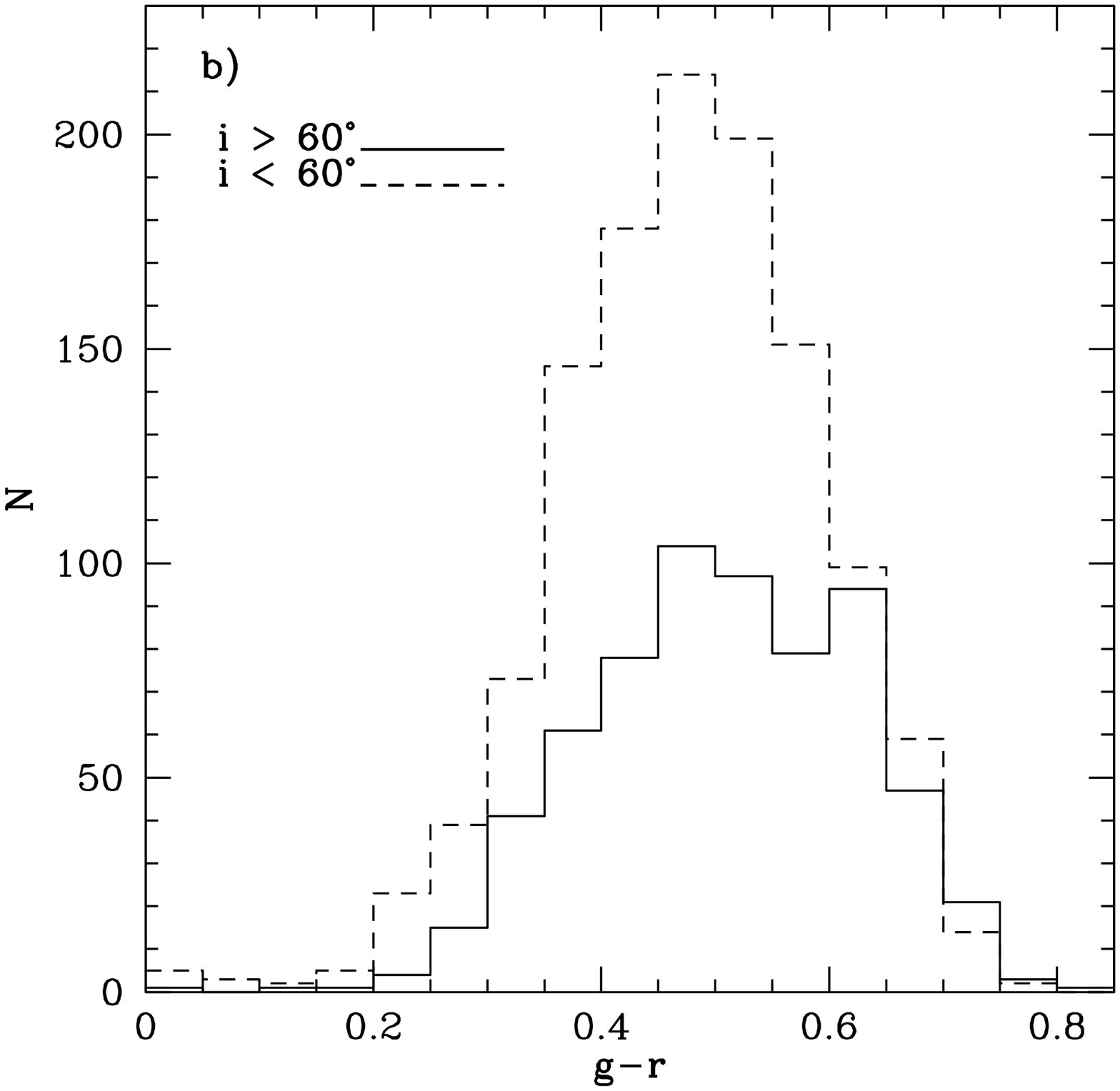}
\caption{{\bf a)} The absolute magnitude distributions of galaxies with
$i>60^{\circ}$ (solid line) compared to the ones with $i<60^{\circ}$ (dashed
line). {\bf b)} The corresponding color distribution, showing the stronger
effect of dust extinction in more inclined disks.\label{incl}}
\end{figure}

\begin{figure}
\epsscale{1}
\plotone{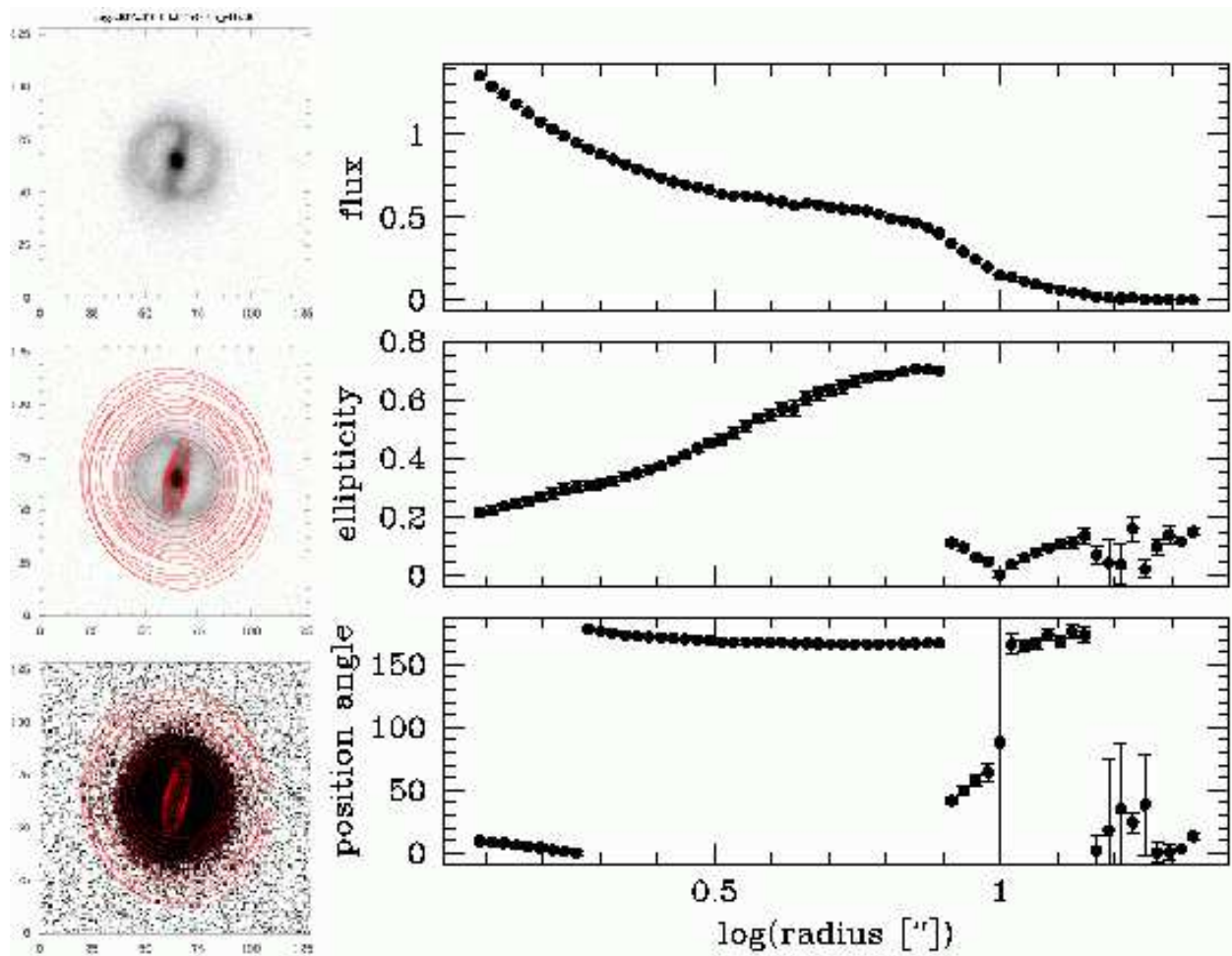}
\caption{The same as Figure \ref{pincl}, but for a galaxy classified as barred from 
the ellipse fits. The images on the left are roughly $50\arcsec$ on a side.
Over the bar region, $e$ rises smoothly to a global maximum of $\sim$~0.7,
while the PA remains $\sim$ constant. After the bar end, as we transition to
the more circular disk, the ellipticity drops sharply at $\sim8\arcsec$ and the
PA changes significantly at this point.
\label{pbar}}
\end{figure}

\begin{figure}
\epsscale{1}
\plotone{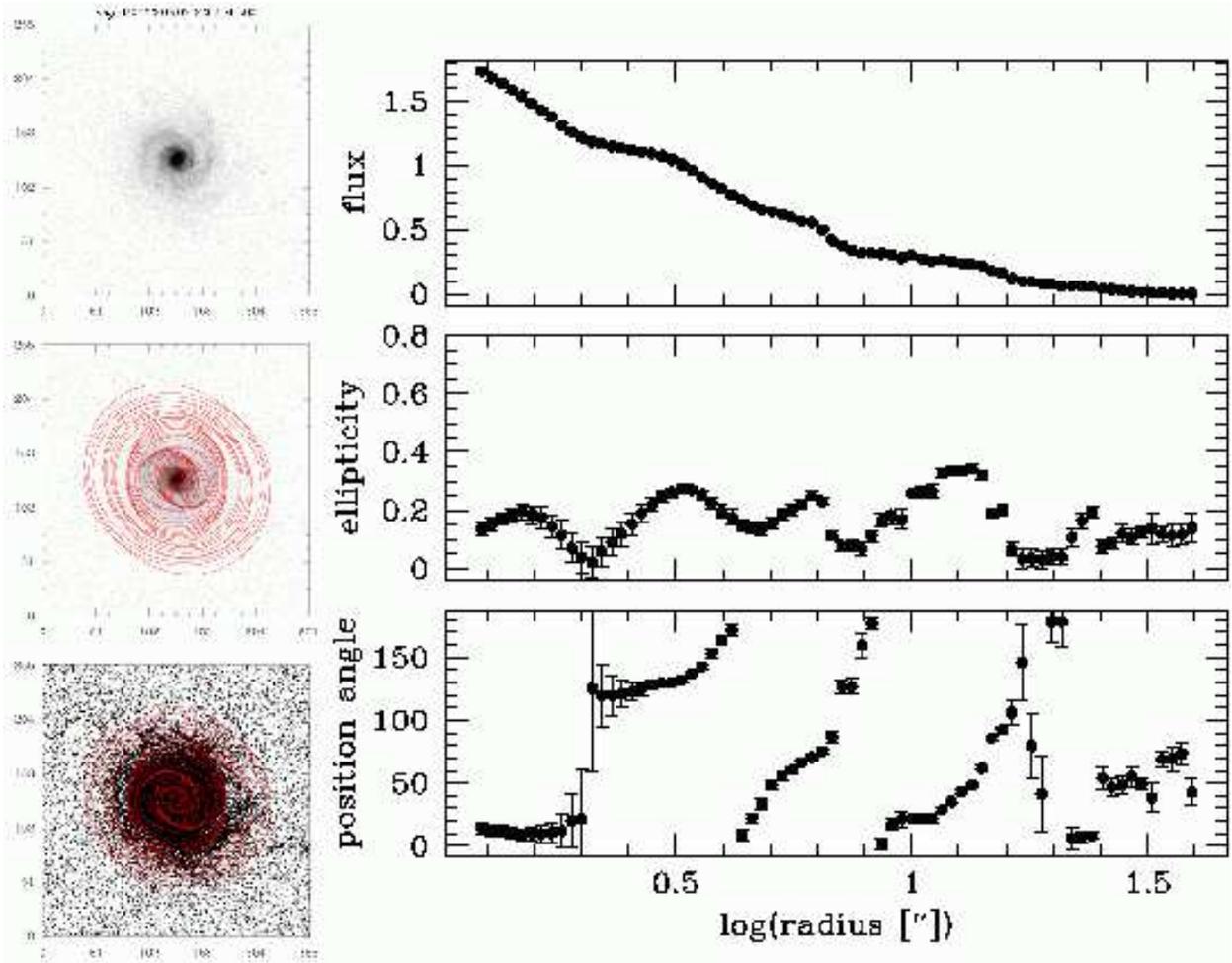}
\caption{The same as Figure \ref{pincl}, but for a galaxy classified as
unbarred. The images on the left are roughly $100\arcsec$ on a side. Here, no
bar signature is evident. Instead the ellipticity profile oscillates and the PA
twists  due to the spiral structure in the disk. 
\label{punbar}}
\end{figure}

When using ellipse fits to characterize bars and disks, we use the method and
steps developed by \cite{jog04} and described in detail in
\cite[][and MJ07]{jog04}. In the following, we give a short description. We use
the standard task 'ellipse' to fit ellipses to the images out to a certain
radius $a_{\rm max}$, which is determined to be the radius where the galaxy
isophotes reach the sky level. An iterative wrapper is used to run 'ellipse' up
to 300 times for each object in order to get a good fit across the whole
galaxy. We performed ellipse fits for the subsample of 1961 disk galaxies
defined by the color cut. For 101 ($\sim5\%$) galaxies, no ellipses could be
fitted, mainly due to the fact that the fitting routine could not find the
galaxy center. This is typically the case for strongly disturbed galaxies, or
on images where a foreground object was not properly removed. In both cases,
the surface brightness of the galaxies is not steadily decreasing from the
center toward larger radii, but is rather oscillating between higher and lower
values impeding a proper ellipse fit. The remaining 1860 galaxies have then
been classified in the way described below. From the best fit for each galaxy,
we plot the associated radial profiles of surface brightness (SB), ellipticity
($e$), and position angle (PA). Furthermore, the fitted ellipses are
overplotted onto the galaxy images to generate overlays. Examples of the radial
profiles and the overlays are shown in Figures \ref{pincl}, \ref{pbar}, and
\ref{punbar}. During the classification process, the plots, the images, and
overlays of the fitted ellipses onto the images, are displayed using an
interactive visualization  tool \citep{jog04}, and used to classify a galaxy as
`inclined', `barred', or `unbarred'. We use the ellipticity in the outer disk
to estimate the inclination $i$. We adopt the standard procedure of excluding
all objects with an inclination $i>60^{\circ}$, as morphological and structural
analysis are unreliable in highly inclined galaxies. Figure \ref{pincl} shows
an example of such a case. We find 648 galaxies ($\sim35\%$) with
$i>60^{\circ}$. In Figure \ref{incl} we show the luminosity and color
distributions of the galaxies with $i>60^{\circ}$ compared to the ones with
$i<60^{\circ}$. The distributions in terms of absolute magnitude are very
similar, whereas the more inclined galaxies tend to be redder than the more
face-on objects. This is expected due to the dust extinction in the disks.

In the next step, we classify the galaxies with $i<60^{\circ}$ as unbarred or
barred, based on the following quantitative criteria: (1) the ellipticity
increases steadily to a global maximum greater than 0.25, while the PA value
remains constant (within $10^{\circ}$), and (2) the ellipticity then drops by
at least 0.1 and the PA changes at the transition from the bar to the disk
region (Figures \ref{pbar} and \ref{punbar} show examples of a barred and
unbarred galaxy, respectively). Criterion (1) is based on the fact that in the
region where bars are dominated by the `$x_{\rm 1}$' family of periodic stellar
orbits \citep{con80}, we expect the ellipses to be aligned along the bar PA and
to become increasingly eccentric toward the end of the bar. Therefore, the
ellipticity should reach a global maximum and the PA should not fluctuate
strongly. This is intuitively evident from the fact that bars appear
morphologically as linear elliptical features centered on the galaxy. The
requirement that the PA must remain constant in the bar region is important for
excluding other features, such as spiral arms that may have a high global
ellipticity. Criterion (2) is applied, because the disks are mostly more
circular than the bar for moderately inclined galaxies and the disk and bar
have different PA in general. After having classified a galaxy, we use the
interactive display tool to measure the ellipticity, PA, and semi-major axis of
its outer disk. For galaxies classified as barred we measure the same
quantities, as well as the maximum ellipticity, $e_{\rm bar}$, of the bar and
the radius, $r_{maxe}$, of maximum bar ellipticity. We use $e_{\rm bar}$ as a
partial measure of the bar strength and the radius $r_{maxe}$ as an estimate
for the semi-major axis of the bar. A detailed theoretical and empirical
justification of this approach is provided in \cite{jog04} and MJ07.

Some of the galaxies exhibit all features required to be classified as barred,
except the `constant PA' criterion, i.e. their PA twists. These galaxies are
classified as `twisted', but regarded as unbarred. However, some of these
galaxies could be weakly barred, since in weak bars the dust lanes on the
leading edge of the bar are curved \citep{ath92b}, which could cause an
isophotal twist. The number of such objects is not very high ($\sim7\%$) and
therefore they do not significantly affect the results.

We note that the classifications and the measurements of sizes, ellipticities
etc. are performed on the {\it observed} images and profiles, which are
affected by projection effects. We did not attempt to deproject our galaxies,
since it is difficult to determine the PA in the outer disks accurately enough,
in order to obtain a reliable deprojection. This is particularly true
for disks with ellipticities $e_{disk}< 0.2$, for which the uncertainty of
the PA determination can reach up to 30--40 degrees in our
data. Furthermore, MJ07 have shown that the statistical results before
and after deprojecting their galaxies are very similar. Finally, we
note that our undeprojected results can be directly compared with
studies of barred galaxies at intermediate redshifts, where 
deprojection has not been carried out \citep{jog04,elm04a,zhe05}.

Out of our sample of 1860 disk galaxies, we find 553 to be barred, 591 to be
unbarred (including 76 classified as twisted), and 648 to be too inclined
($i>60^{\circ}$). We did not classify the remaining 68 galaxies for the
following reasons. For 30 galaxies, the ellipse fits obviously failed or the
profiles were extremely messy: this occurred, when foreground/background
objects are not completely removed or for galaxies, with very low surface
brightnesses and very irregular shapes. The other 38 galaxies were ambiguous
cases where deprojection might make a large difference. Specifically, in these
galaxies, the ellipticity rises smoothly to a {\it local} maximum while the PA
stays relatively constant, but this maximum ellipticity is less than that of
the outer disk. In such cases, deprojection may turn the local maximum into a
global maximum, particularly if the bar is perpendicular to the line of nodes.
These objects are further discussed in $\S$ \ref{resol}.

In Table \ref{basic} we give the median values of the basic properties for the
barred, unbarred, inclined, and unclassifiable objects.

\section{Detectability of bars}\label{resol}

When using optical images, the obscuration of bars by dust and star formation 
can prevent their unambiguous detection. Comparing the results of
quantitative bar studies conducted in the optical and in the
near-infrared, shows that the NIR bar fraction can be a factor of
$\sim1.3$ higher than in the optical \citep[e.g.][]{mar07,men07}. It is
therefore clear that we miss bars due to dust extinction and that our
bar fraction has to be considered as a lower limit. Furthermore, our
results can only be compared to studies, which have also been
performed using observations in the optical.

Apart from dust, the ability to detect a bar in a galaxy depends on
its distance, its inclination, the angle $\theta$ between the line of
nodes and the PA of the bar, as well as on the point spread function
(PSF) or seeing of the images used. We discuss each of these factors
below. The largest allowed inclination in our sample is
$i$~=~$60^{\circ}$. When computing the smallest measurable bar
diameter $d_{\rm min}$, we assume that a bar is detectable only if its diameter
can encompass at least 2.5 times the PSF ($1\farcs4$) of the images. The median
seeing or PSF of the $r$-band images from SDSS is $1\farcs4$, which corresponds
to $\sim290$ pc at $z=0.01$ (our lower redshift limit) and to $\sim840$ pc at
$z=0.03$ (our upper redshift limit). If a bar, which would be detectable in a
face-on disk, happens to be perpendicular to the line of nodes in an inclined
disk (i.e. $\theta=90^{\circ}$), projection effects will reduce its apparent
length and ellipticity compared to the intrinsic value, and this length may
fall below the detection limit. For values of $\theta$ below $90^{\circ}$,
these effects are less severe.

\begin{figure}
\epsscale{1}
\plotone{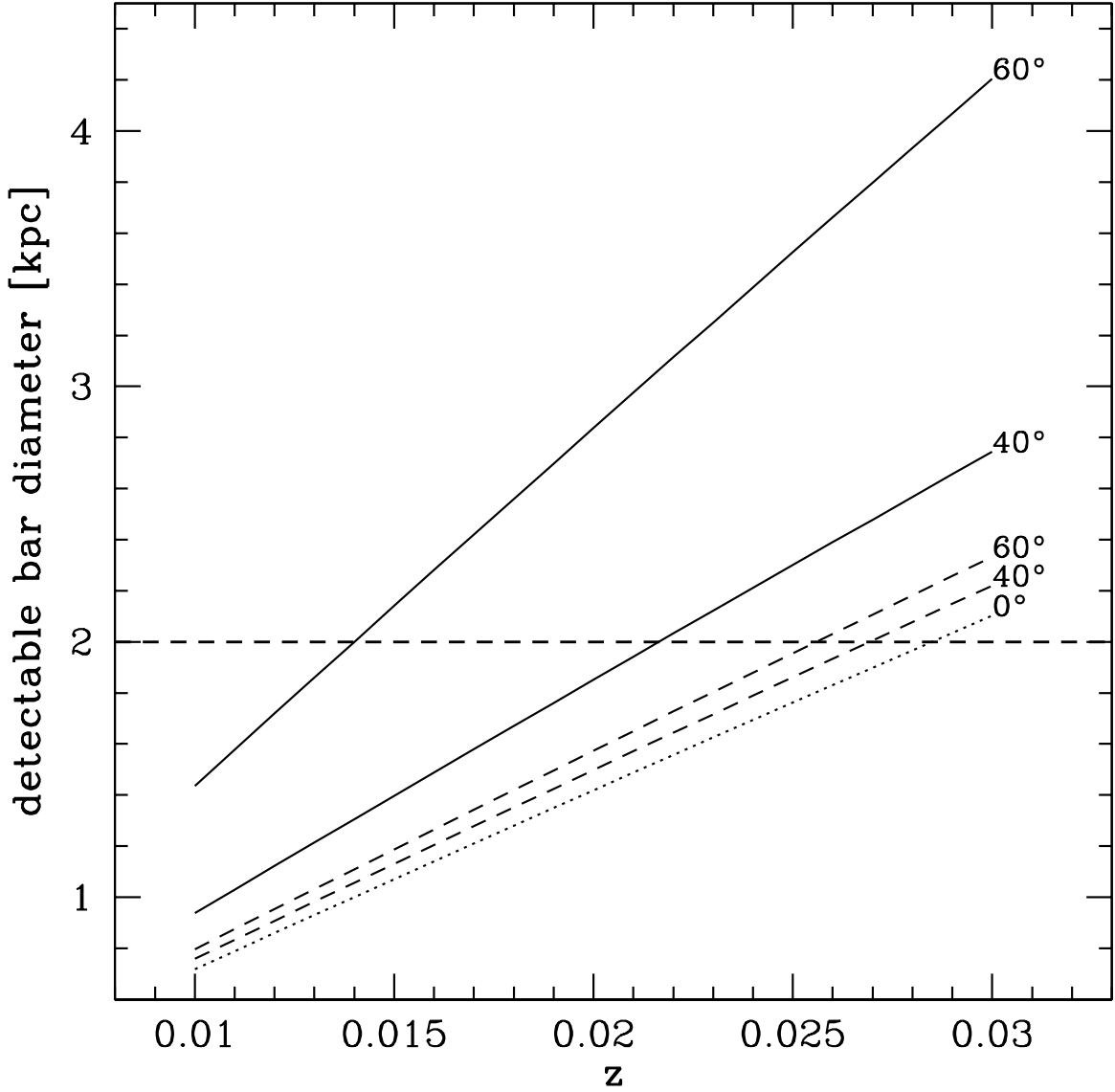}
\caption{The horizontal line corresponds to the minimum bar diameter (2 kpc) of
large-scale bars. The 5 diagonal lines show the smallest measurable bar
diameter $d_{\rm min}$ as a function of redshift, for different values of the 
angle $\theta$ between the bar PA and the galaxy's lines of nodes, and for
different inclinations $i$. When computing $d_{\rm min}$, we assume that a bar
is detectable only if its diameter can encompass at least 2.5 times the PSF
($1\farcs4$) of the images. The 2 solid lines show the detection limit for the
worst case scenario of $\theta$~=~$90^{\circ}$, and two inclinations
($i$~=~$40^{\circ}$ and $60^{\circ}$). The 2 dashed lines show the detection
limit for a more moderate $\theta$~=~$30^{\circ}$. The dotted lines shows the
detection limit for the face-on case, where independent of $\theta$, bars with 
diameters  $\gtrsim 2$ kpc are detectable out to $z=0.03$.
\label{see}}
\end{figure}

The variation of the detection limit $d_{\rm min}$ as a function of $\theta$,
$i$, and $z$ is shown by the 5 diagonal lines in Figure \ref{see}. The
horizontal line  corresponds to the minimum bar diameter (2 kpc) of large-scale
bars, and ideally, we want the detection limit to lie below this line at all
redshifts. The two solid diagonal lines indicate the detection limits as a
function of redshift for two different inclinations ($i$~=~$40^{\circ}$ and
$i$~=~$60^{\circ}$) and the worst case scenario of $\theta$~=~$90^{\circ}$. In
that case, with  $i$~=~$40^{\circ}$ and $\theta$~=~$90^{\circ}$, we start
missing the smallest large-scale bars at $z >$~0.02. The two dashed diagonal
lines on Figure \ref{see} represent the detection limits for a more moderate
$\theta$ of $30^{\circ}$. In that case, with $i$~=~$40^{\circ}$ and
$\theta$~=~$30^{\circ}$, we start missing the smallest large-scale bars at
$z>$~0.027. Finally, the dotted diagonal line on Figure \ref{see} represents
the  face-on case ($i$~=~$0^{\circ}$) where, independent of $\theta$, bars with
diameters $\gtrsim 2$ kpc are detectable out to $z=0.03$.

\begin{figure}
\epsscale{1}
\plotone{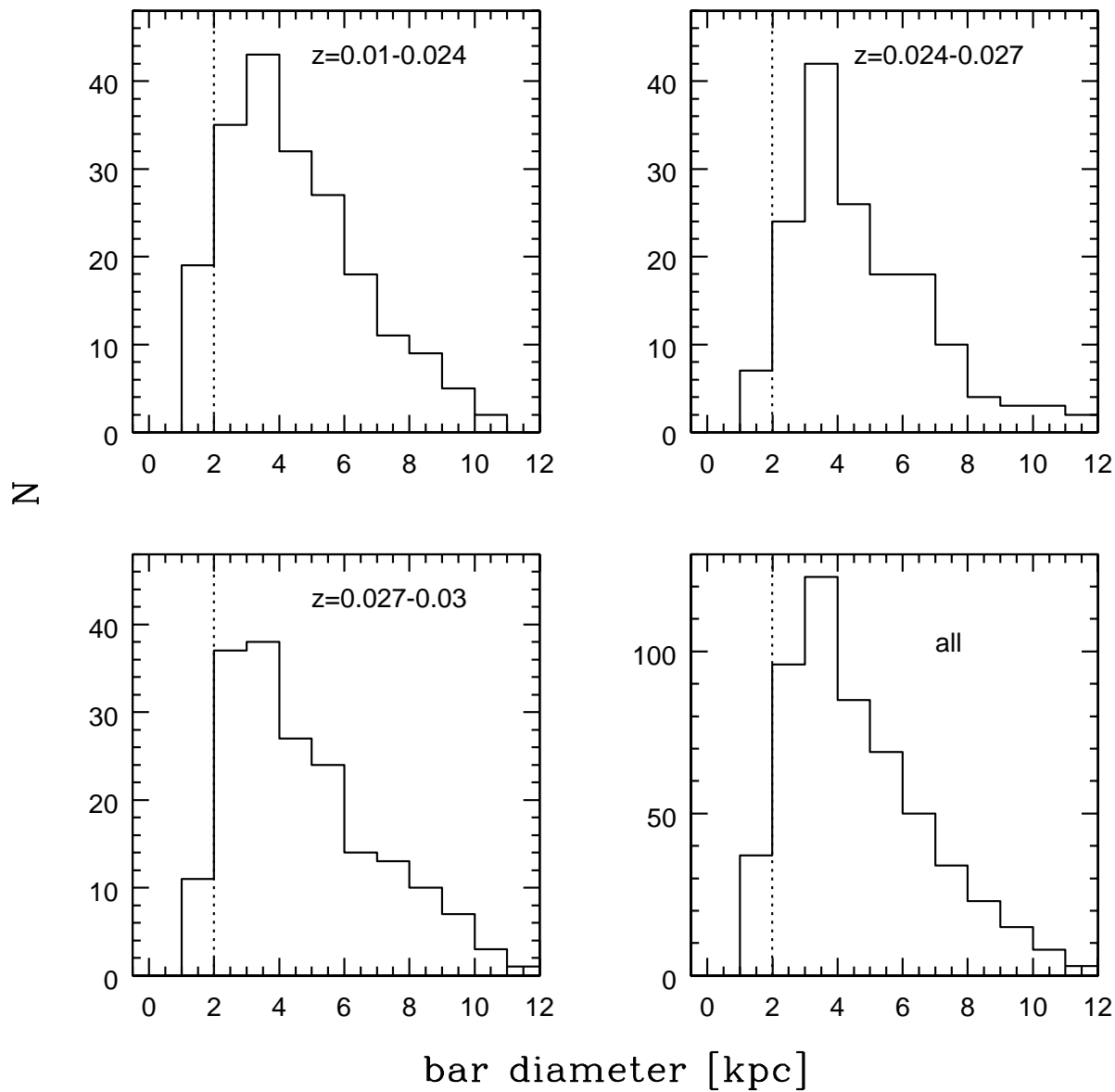}
\caption{The distribution of the bar diameters in three different redshift bins
and for the total sample (the bin sizes have been chosen in order to obtain
roughly the same number of objects in each bin). The vertical lines indicate
the lower limit of 2 kpc for diameters of large-scale bars. The distributions are very
similar, in particular we do not miss small bars at higher redshifts.
\label{red}}
\end{figure}

In Figure \ref{red}, we plot histograms of the absolute bar diameters in three
different redshift bins and for the whole sample (the bin boundaries have been
chosen in order to obtain roughly the same number of objects per bin). The
vertical lines indicate our bar diameter limit of 2 kpc. Figure \ref{red} shows
that we do not miss significant numbers of small bars with increasing redshift.
We find roughly the same number of bars with $2<d_{\rm min}<3$ kpc in the
lowest and in the highest redshift bins.

\begin{figure}
%\epsscale{2.3}
\plottwo{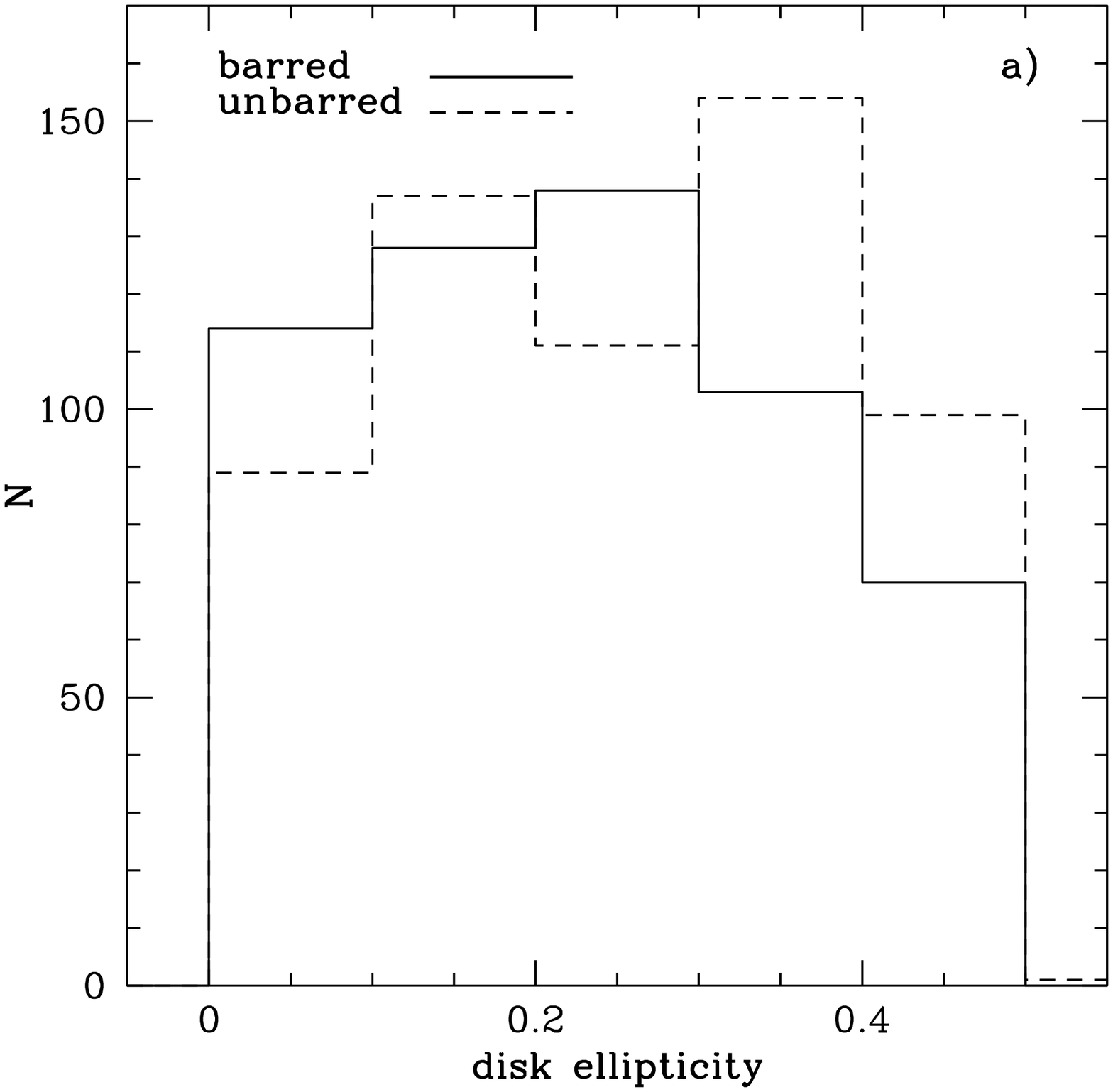}{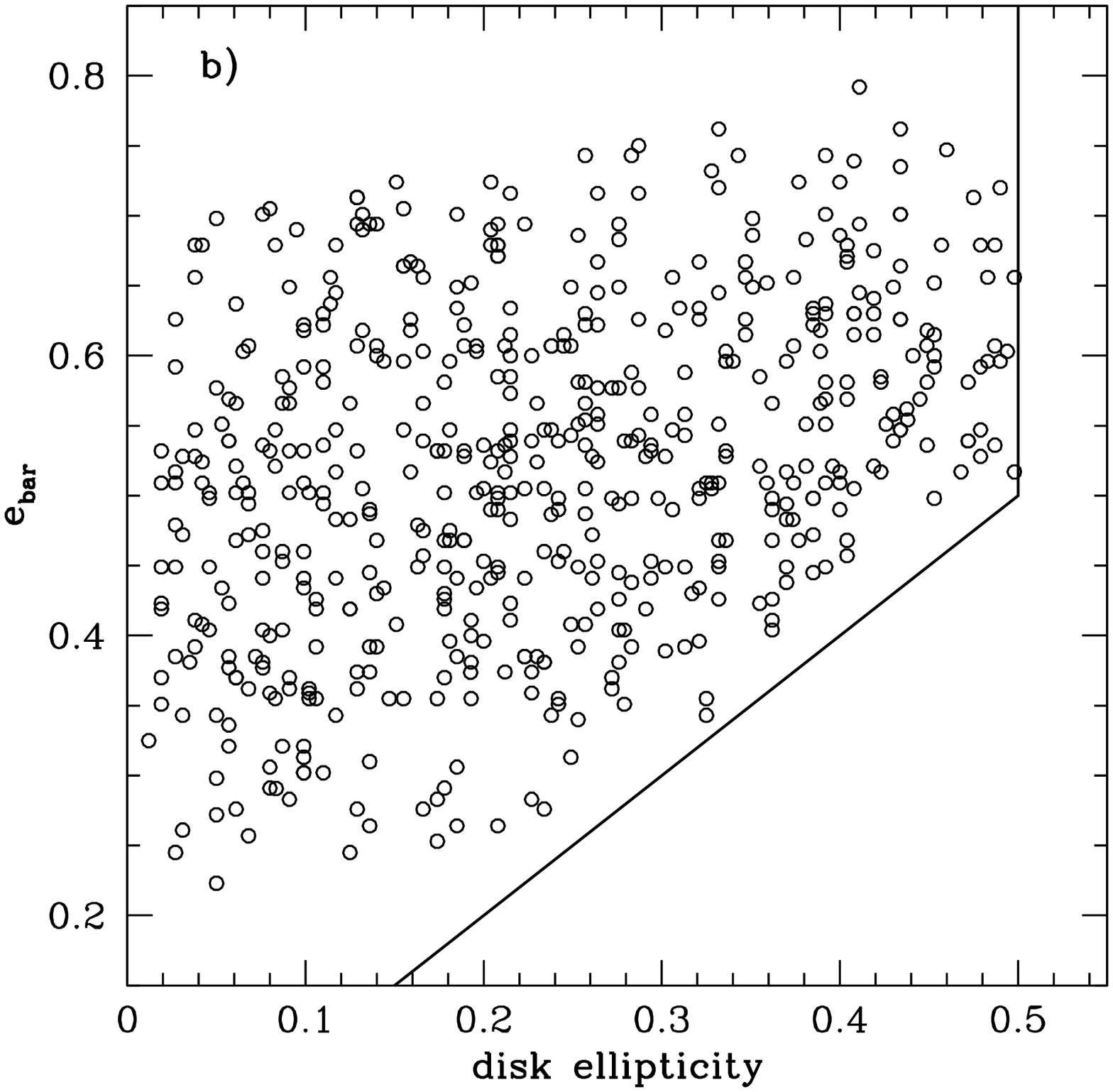}
\caption{{\bf a)} The disk ellipticities ($e_{\rm disk}$) of the barred (solid
line) and unbarred (dashed line) subsamples. As $e_{\rm disk}$ varies from 0.0
to 0.5, the total number of barred objects decreases only slightly at $e_{\rm
disk} \gtrsim0.3$. This fall can be attributed to projection effects caused by 
the inclination of the disk and our criterion (1) for bar detection (see text
for details). {\bf b)} Plot of the disk ellipticity versus bar ellipticity
($e_{\rm bar}$) for galaxies classified as barred. The vertical
line at  $e_{\rm disk}$ of 0.5 reflects the fact that all highly inclined
($i>60^{\circ}$) disks were excluded from the sample in order to ensure
reliable morphological analyzes. The diagonal solid line is defined by $e_{\rm
disk}$~=~$e_{\rm bar}$. All detected bars lie to the left of this line,
reflecting the criterion (1) that the bar ellipticity $e_{\rm bar}$ must be a
global maximum. We note that maximum $e_{\rm bar}$ is similar at different
$e_{\rm disk}$, indicating that the detection of strong bars is not biased to
the more inclined disks.
\label{ell}}
\end{figure}

In order to gauge the impact of projection effects on the derived bar
ellipticities, we plot the distributions of barred and unbarred galaxies as a
function of disk ellipticity in Figure \ref{ell}a. The cutoff at disk
ellipticities ($e_{\rm disk}$) at 0.5 is due to the fact that highly inclined
disks with $i > 60^{\circ}$  are discarded from our sample ($\S$ \ref{meth}). 
As $e_{\rm disk}$ varies from 0 to 0.5, the total number of barred objects 
decreases only slightly at $e_{\rm disk} \gtrsim0.3$. This fall could be
attributed to two related factors. For a feature to be classified as a bar, our
criterion (1) in $\S$ \ref{meth} requires its maximum ellipticity to reach a
{\it global} maximum, which is higher than the ellipticity  $e_{\rm disk}$ of
the outer disk. At high inclinations, where $e_{\rm disk}$ is high, this
criterion may not be satisfied even if the feature is truly a bar. Some of the
38 `unclassifiable' galaxies that we discussed at the end of $\S$ \ref{meth}
fall in this category. A second factor is that if $\theta$ is close to
$90^{\circ}$, then at high inclination $i$ or $e_{\rm disk}$, the bar
ellipticity is lowered  significantly, making it fail criterion (1). However,
it is clear from Figure \ref{ell}a, that the number of bars we might miss in
the two highest ellipticity bins ($\epsilon\gtrsim0.3$) is not  very high. 

The distribution of  bar  ellipticities ($e_{\rm bar}$) in disks of different
ellipticities ($e_{\rm disk}$) is shown in Figure \ref{ell}b. The vertical line
at $e_{\rm disk}=0.5$ indicates the exclusion of highly inclined disks. The
diagonal solid line is defined by  $e_{\rm disk}$~=~$e_{\rm bar}$. All detected
bars lie to the left of this line, reflecting the criterion (1) that the bar
ellipticity $e_{\rm bar}$ must be a global maximum. We note that the maximum
$e_{\rm bar}$ is similar at different $e_{\rm disk}$, indicating that the
detection of strong bars is not biased to the more inclined disks.

In the next sections, we perform checks to verify that the results presented
are not dominated  by the afore-discussed detection limits. We verify, for
instance, that the results over $0.01\leq z<0.03$, also hold in the lower
redshift bins $0.01\leq z<0.02$. We also verify that derived properties of bars
hold for bars with a wide range of diameters, and are not biased by the
smallest large-scale bars with  diameters close or slightly higher 2 kpc, since
the latter bars are more susceptible to detection problems.

\section{Results}\label{resul}
\subsection{The globally-averaged optical bar fraction and bar properties at
$z\sim0$}\label{glob}
As noted in $\S$ \ref{samsel}, our sample includes both bright early-type
galaxies with bulges, and many disk-dominated galaxies of late Hubble types,
while samples in earlier studies
\citep[e.g.,][MJ07]{esk00,elm02,lau05,lau06,but06} were dominated by the former
group of galaxies. In $\S$ \ref{bfp}, we will investigate the bar properties as
a function of different galaxy types, but for now, we begin by estimating
globally-averaged properties across our full sample.

The optical $r$-band bar fraction is defined as the fraction of disk galaxies
with $i<60^{\circ}$ that host large scale bars, measured on optical images. We
find 553 bars among 1144 moderately inclined disk galaxies. Hence, the optical
$r$-band bar fraction, averaged over our sample, which is biased
toward late-type disk dominated galaxies, is $\sim48\%$. This number
could be a lower limit, due to missing weak bars with isophotal
twists. As discussed above we found 76 objects with these
characteristics, which represent $\sim7\%$ of the sample. Our result
is in good agreement with \cite{agu07}, who found an optical $r$-band
fraction of $\sim45\%$, which is also based on SDSS. As stated in
$\S$ \ref{intro}, the optical bar fraction is a lower limit to the
total bar fraction as the optical images miss bars obscured by
dust and star formation. Such bars can be detected in the near-infrared and
recent studies show that the near-infrared bar fraction is $\sim15\%$ higher
than the optical bar fraction, or around $\sim$~60\% \citep[MJ07,][]{lau04,men07}.

Figure \ref{barp}a shows the distribution of bar semi-major axis lengths for
the 553 bars in our sample. We find 37 bars with a semi-major axis $<1$ kpc,
which are commonly considered as nuclear bars, and the remaining are 516
large-scale bars. We exclude the 37 nuclear bars in the following plots and
discussions, which affects the optical bar fraction by only $\sim1\%$. Most
large scale bars ($\sim72\%$) have sizes in the range 1 to 3 kpc (Figure 
\ref{barp}a). We find relatively few bars with sizes $>5$ kpc. The median bar
size of the sample is 2.2 kpc.

The distributions of bar ellipticities is shown in Figure \ref{barp}c. The
majority of bars have ellipticities in the range 0.3 to 0.7. In general, the
distribution is in good agreement with the corresponding $H$-band results of
MJ07. We do not find any strong relation between bar ellipticity and absolute
magnitude (Figure \ref{barp}d), or between bar length and absolute magnitude
(Figure \ref{barp}b). However, as discussed in $\S$~5.4, the maximum
ellipticity or bar strength is on average higher in faint quasi-bulgeless
galaxies than in galaxies with bulges (see Figure \ref{visp2}d).

In order to be able to normalize the bar size, we determine the isophotal
radius, at which the surface brightness reaches 24 mag/arcsec$^2$ ($R_{24}$).
Since we are using $r$-band images, this radius corresponds roughly to
$R_{25}$, the radius, where the $B$-band surface brightness reaches 25
mag/arcsec$^2$. In Figure \ref{r24}a we plot $R_{24}$ versus the bar size.
There is no strong correlation between these two parameters, however, the bar
length is typically much smaller than $R_{24}$. Figure \ref{r24}b shows the
ratio $a_{bar}/R_{24}$ versus the absolute $g$-band magnitude. Most objects
have values in the range 0.2 to 0.4 (median 0.32), which is consistent with the
results of MJ07 and \cite{erw05}. Theory predicts that the bar ends between the
4:1 and the corotation resonance (CR). Furthermore, studies of the pattern
speeds of bars suggest that the bar ends very near the CR, as they find that
the ratios between the bar length and the CR are in the range 0.7 to 0.9
\citep{agu03,deb04}. If these values are representative, then our result that
$a_{bar}/R_{24}$ is primarily well below 1, suggests that the CR lies well
inside $R_{24}$ in most disks.

\subsection{The optical bar fraction as a function of half-light radius, 
$r_{\rm e}/R_{24}$, luminosity, and color}\label{bfp}

The large number of disk galaxies in our sample allows us to perform for the
first time a statistically significant study of the dependence of the bar
fraction on other galaxy properties, such as  luminosity, measures of
bulge-to-disk ($B/D$) ratios, size, colors, surface brightness, etc. In effect,
we can bin the data as a function of different parameters with up to 100
galaxies in each bin. We note that this was not possible in earlier studies,
which have total sample sizes of $\le250$ galaxies. Our study can therefore
help to understand, which galaxies are more likely to form or maintain a bar
and how the presence of a bar relates to the evolution of a galaxy.

For all the results presented in this paper, we omit objects with half light
radii $r_{\rm e}<2$ kpc. As we will show in $\S$ \ref{visual}, this regime is
strongly contaminated by non-disk galaxies, such as dwarf spheroidals. Without
these galaxies the optical $r$-band fraction is $52\%$. In Figure \ref{barf} we
show the optical $r$-band bar fraction ($f_{\rm opt-r}$) as a function of
specific galaxy properties. In all panels the numbers next to the points
indicate the total number of objects in the corresponding bins. The dashed
lines indicate the total optical bar fraction ($52\%$). We only plot bins with
more than 10 objects. 

Figure \ref{barf}a shows the optical $r$-band bar fraction as a function of the
half-light radius ($r_{\rm e}$) derived from the single-component 2D S\'ersic fit
($\S$ \ref{samsel}). The radius $r_{\rm e}$ encloses half of the total light of the
galaxy and is a measure of the central light concentration of the galaxy. 
Figure \ref{barf}a shows that the  optical bar fraction rises steadily, from
$\sim$~40\%--50\% in galaxies with small $r_{\rm e}$ (2--3 kpc) to $\sim$~60\% for
galaxies with large  $r_{\rm e}$ ($\gtrsim4$ kpc). At a given luminosity, galaxies
with a large bulge-to-disk ($B/D$) ratio typically have a smaller $r_{\rm e}$ than
disk-dominated galaxies with no bulge or only  a very low $B/D$. At this point,
one may be tempted to ask whether the drop in the optical $r$-band bar fraction
in Figure \ref{barf}a, as $r_{\rm e}$ drops from 4 to 2 kpc, is due to the bar
being too small to be detected. This is not the case because {\it $r_{\rm e}$ is
not equivalent to the size of the disk or bar component}. In particular,
galaxies with a large $B/D$ and low $r_{\rm e}$ may have an extended disk and a
large bar. This is shown in Figures \ref{bsize}a and \ref{bsize}b, where
$r_{\rm e}$ is plotted against the bar semi major axis $a_{\rm bar}$ for 2 redshift
bins. It is evident that for $r_{\rm e} \sim$~2 to 4 kpc, $a_{\rm bar}$ ranges from
1 to 5 kpc, and is easily resolved. Thus, the trend in optical bar fraction
with $r_{\rm e}$ in Figure \ref{barf}a seems to be a solid one. We further
investigate this below using $r_{\rm e}/R_{\rm 24}$.
 
In Figure \ref{barf}b, we plot the optical bar fraction as a function of
the normalized $r_{\rm e}$, using $R_{\rm 24}$ as normalization.  At a given
luminosity, $r_{\rm e}/R_{\rm 24}$ is a measure of the relative light distribution
in the bulge and disk, and hence a rough measure of the $B/D$ light ratio. The
ratio $r_{\rm e}/R_{24}$ is not correlated to $M_g$ (not shown). Figure \ref{barf}b
is very similar to Figure \ref{barf}a, also showing a steep increase of the bar
fraction towards larger, more extended galaxies. In fact, the effect is more
pronounced in Figure \ref{barf}b, where the optical bar fraction reaches only
$\sim30\%$ for the most compact disks and rises to more than $60\%$ for the
most extended disks. This shows that the increase in bar fraction is not
primarily a luminosity effect, but is related to the structure of the disk.
Therefore, Figure \ref{barf}b  can be interpreted as indicating that the {\it
optical bar fraction is higher in disk-dominated systems}. Visual
classification, as discussed in detail in $\S$ \ref{visual}, confirms this
interpretation.

In Figure \ref{barf}c, $f_{\rm opt-r}$ is roughly constant at $M_g<-19.5$ mag
(neglecting the brightest bin, which is very small) and increases towards the
fainter end of the magnitude range, reaching almost $60\%$ for the faintest
bin. This is consistent with the above interpretation of a higher bar fraction
in disk-dominated galaxies, since the latter dominate at fainter magnitudes.

In Figure \ref{barf}d we plot the optical $r$-band bar fraction as a function
of $g-r$ color. Notice the sharp increase in bar fraction as the $g-r$ color
gets bluer from 0.55 to 0.30.  There are 2 potential interpretations of this
trend. One interpretation is that star-forming galaxies host an excess of bars
\citep[e.g.][]{hun99} as the star formation is bar-induced. However, looking at
the $g$-band images of our sample galaxies, we do not find that barred
late-type disks show more centrally concentrated star formation than unbarred
galaxies. Another more likely interpretation is that the higher optical bar
fraction in late-type, disk-dominated galaxies, suggested by Figure
\ref{barf}b, naturally leads to a higher optical bar fraction for bluer colors,
because late-type galaxies tend to be bluer and have higher specific star
formation rates \citep{gav98,ben02,koo04}. The analysis in $\S$~5.6 and Figure
\ref{split} further support this interpretation.

The fact that the relationship between optical bar fraction and blue
colors has not been reported in earlier studies \citep[MJ07,][]{esk00} is
likely due to the fact that their samples were dominated by brighter
earlier-type galaxies, while ours has a large number of late-type
disk-dominated galaxies ($\S$ \ref{samsel}).

\subsection {The optical bar fraction as a function of $n$ and
$\mu_0$}\label{ertest}
In Figure \ref{normfcsb}a, we plot $f_{\rm opt-r}$ as a function of the
S\'ersic index $n$. The low $f_{\rm opt-r}$ for $n>$~2.5 are likely due to blue
spheroids contaminating the color-selected sample of disk galaxies (see also
Figure \ref{cuts}). The rise in $f_{\rm opt-r}$ at $n\le$~1.5 is consistent
with a larger bar fraction in disk-dominated systems. 

In order to further investigate the assumption that the bar fraction is related
to the presence and size of a bulge, we measure the central surface
brightnesses ($\mu_0$) of the galaxies directly on the $r$-band images. This is
an important test as the measurement of $\mu_0$ provides a measure of the
central light concentration, which is {\it model-independent} and not affected
by the bar itself. In contrast, the half light radius $r_{\rm e}$ was derived from
a S\'ersic fit, and it is possible that  parameters, such as $r_{\rm e}$ or
$R_{24}$, are affected by the  details of the fit or the presence of a bar. For
instance, bars dominating the light distribution in disks could automatically
lead to larger $r_{\rm e}$ and $R_{24}$, because they efficiently disperse the
luminosity. This would, however, only be the case in galaxies, where the bar is
much more luminous than the bulge and the underlying disk. We measured $\mu_0$
on the same physical size (1 kpc$^2$) for all galaxies. In Figure
\ref{normfcsb}b, we plot the optical bar fraction as a function of $\mu_0$. The
bar fraction is steadily rising for decreasing $\mu_0$. This result is similar
to the ones for $r_{\rm e}$ and $r_{\rm e}/R_{24}$. However, the change of the bar
fraction with respect to $\mu_0$ is less steep, but more continuous. This
result lends support to the view that bars are more likely to be found in disks
with relatively small  bulges.

\subsection{Visual classifications of disk-dominated versus early-type
galaxies}\label{visual}
In $\S$ \ref{bfp} and \ref{ertest}, we interpreted the higher optical bar
fraction at larger $r_{\rm e}$ (Figure \ref{barf}a) and $r_{\rm e}$/$R_{\rm 24}$
(Figure \ref{barf}b) as meaning that disk-dominated galaxies with very low
bulge-to-disk ($B/D$) ratio have a higher optical bar fraction. The most
rigorous way to test this claim is to perform 3 component bulge+bar+disk
decomposition on the 2D light distribution of the galaxy, and derive a $B/D$.
However, this task is beyond the scope of the present paper. Instead, we
perform  a first order test by visually classifying $\sim80\%$ of our sample,
which includes galaxies with $r_{\rm e}$ in the range 2 to 10 kpc. Our main goal in
this visual classification is to identify late-type disk-dominated galaxies
with no significant bulge, as well as early-type galaxies with bulge and disk
components, so that we can compare their bar fractions. We also classify the
sub-group of systems with $r_{\rm e}<2$ kpc, which we excluded from our sample in
$\S$ \ref{bfp}, on the ground that this group is strongly contaminated by pure
bulge/spheroidal galaxies, such as dwarf spheroidals or dwarf ellipticals. A
secondary goal of the visual classification is to identify pure bulge systems
and verify that they indeed cluster at $r_{\rm e}<2$ kpc.

Visual inspection does not allow one to classify galaxies in fine grids of
$B/D$ ratios, but it does allow one to reliably classify galaxies into three
broad visual classes (VCs):
\vspace{-4mm}
\begin{description}
\vspace{-3mm}
\item[Class 1:] pure bulge or spheroid: steady decrease of the surface
brightness from the center outward, with no obvious break in the surface
brightness profile
\vspace{-3mm}
\item[Class 2:] disk galaxy with bulge
\vspace{-3mm}
\item[Class 3:] pure disk with no bulge: no bright, distinct, and roughly round
central object
\end{description}

The classification has been performed by all three authors, making sure that
each object is classified by at least two classifiers. A small fraction of the
objects have been classified twice to test the robustness of the results. The
agreement between classifiers is very good ($>80\%$) and the final result
represents an average of all classifications. In Figure \ref{ima} we show
images of 16 representative objects. The galaxies in the first row have been
classified as pure spheroids (class 1), objects in row two and three are in
class 2, and the fourth row shows examples of pure disks (class 3). 

The first result of our visual classification is that our sample of
1144 disk galaxies (with $i<60^{\circ}$), which was color selected
($\S$ \ref{samsel}), only has a small contamination of $\sim7\%$ by
pure spheroids (class 1). Many of these objects
in class 1 are relatively blue and faint. These objects could be dwarf
ellipticals, which experienced a recent episode of star formation and whose
luminosity is therefore dominated by rather blue stars. It is also interesting
to note that almost all of these objects have $n<2.5$ and would also have been
included in our sample of disk galaxies if we had used a S\'ersic cut to select
the sample (e.g., see Figure \ref{cuts}). The main point to note is that the
small fraction ($\sim7\%$) of class 1 objects would not have any significant 
impact on our globally-averaged results, such as the global optical bar
fraction ($\S$ \ref{glob}). However, the contamination particularly affects the
results in the lowest $r_{\rm e}<2$ kpc bin as $\sim95\%$ of the class 1 (pure
spheroids) objects are very compact and fall in this bin. This results in a
spheroid contamination of up to $20\%$ for objects with $r_{\rm e}<2$ kpc, which
makes the bar fraction for this subsample very uncertain. Thus, throughout this
paper ($\S$ \ref{bfp} onward), we excluded all objects with $r_{\rm e}<2$ kpc from
the plots and subsequent analysis. We estimate that the contamination by
spheroids for the remaining sample is $<1\%$.

Next, we discuss the galaxies that are visually classified as class 2
(bulge+disk) and class 3 (bulgeless) galaxies. In the sample of 886 galaxies
with $r_{\rm e}>2$ kpc, $\sim20\%$ of galaxies fall in class 3 (bulgeless), while the
remaining fall in class 2. It is  remarkable that {\it the optical bar fraction
of the class 3 (bulgeless) disk galaxies is $\sim87\%$, compared to $\sim44\%$
for class 2 (bulge+disk) galaxies.} This striking difference supports the basic
conclusion of $\S$ \ref{bfp} and \ref{ertest}: {\it disk-dominated galaxies
with no bulge or a very low $B/D$ display a much higher optical bar fraction
than galaxies with significant bulges}. In fact, it appears that a pure disk is
twice as likely to be barred than a disk galaxy with a bulge. We note that the
higher bar fraction found for class 3 objects is also consistent with the high
bar fractions in late-type galaxies  reported by \cite{ode96} and \cite{elm04a}
based on RC3 visual bar classes and RC3 Hubble types.

Another way to illustrate the results is to look at the difference in disk
properties between barred and unbarred galaxies. Figure \ref{visp} shows the
percentage of class 3 (bulgeless) and class 2 (bulge+disk) galaxies among
barred galaxies (solid histograms) and unbarred galaxies (dashed histogram).
The fraction of bulgeless galaxies in much higher ($31\%$ vs $5\%$) in barred
than unbarred systems. Figures \ref{visp2}a and b show the distributions of
$r_{\rm e}$/$R_{\rm 24}$ among barred galaxies (solid histograms) and unbarred
galaxies (dashed histograms) for galaxies brighter than the median luminosity
of the sample (a) and for galaxies fainter than the median (b). The fraction of
galaxies with large $r_{\rm e}$/$R_{\rm 24}$ ratios is higher in barred than
unbarred systems, particularly for the fainter subsample. Figures \ref{visp2}c
and d show the distributions of bar ellipticities for galaxies in class 3 (solid
histograms) and galaxies in class 2 (dashed histograms), again for the bright
and faint subsamples. These figures indicate that the bars in bulge-dominated
galaxies are generally weaker than in disk-dominated galaxies. It seems that
the presence of a bulge weakens the bar, in particular in fainter galaxies
(panel d). However, one has to keep in mind that $e_{\rm bar}$ has been
determined including the bulges and that the measured bar ellipticities may be
affected by the bulges, in the cases where the end of the bar is close to
the bulge.

In summary, the visual classifications have provided two important results.
They show that the contamination from non-disk galaxies is small and confined
to the regime $r_{\rm e}<2$ kpc. These galaxies do not impact our result as the
regime $r_{\rm e}<2$ kpc is excluded from all analysis in this paper. Secondly, and
more importantly, the visual classifications support the claim, made in $\S$
\ref{bfp} and \ref{ertest}, that disk-dominated galaxies with a very low $B/D$
display a significantly higher optical bar fraction (60\% to 70\%) than
galaxies with a significant bulge (40\% to 50\%). The  associated 
ramifications are discussed  in $\S$ \ref{discu}.
 
\subsection{The bar fraction as a function of mass}\label{fmass}
The mass of galaxy disks is one of the fundamental parameters controlling their
evolution. We use the prescription of \cite{bel03} to derive stellar
mass-to-light ratios using the $g-r$ color:
\begin{displaymath}
\log(M/L_r)=-0.306+1.097(g-r)
\end{displaymath}
This mass-to-light ratio is then used to derive the stellar masses of the
galaxies using the following relation:
\begin{displaymath}
\log(M)=\log(M/L_r)-0.4(M_r-r_{\odot})
\end{displaymath}
where $r_{\odot}=4.67$ is the absolute $r$-band magnitude of the Sun. 

The optical bar fraction as a function mass is shown in  Figure \ref{massp}. 
Over the mass range $5 \times 10^{9}$ to $5 \times 10^{10}$ $M_{\odot}$, the
optical bar fraction rises for lower masses. This trend is expected from our
earlier findings ($\S$ \ref{bfp} and \ref{ertest}) of a higher optical bar
fraction in galaxies, which are more disk-dominated, less centrally
concentrated, bluer, and fainter.  

\subsection{Which disk parameters most strongly influence the optical bar
fraction?}\label{psplit}

In $\S$ \ref{bfp} to \ref{fmass}, we showed that the optical bar fraction
rises with lower bulge-to-disk ratios, as characterized visually (Figure \ref{visp}) 
and also via  $r_{\rm e}$/$R_{\rm 24}$ (Figure \ref{barf}b); with lower 
central surface brightness (Figure \ref{normfcsb}b), and with bluer
$g-r$ colors (Figure \ref{barf}d). There is also a weaker trend with
fainter absolute magnitude $M_{\rm g}$ (Figure \ref{barf}c) and with
lower masses (Figure \ref{massp}).

Many of these parameters are correlated and the above findings are all 
consistent with the optical bar fraction rising toward late type  
galaxies (e.g., Sd,Sm). The latter systems have no bulge or very low
bulge-to-disk ratios, low central mass concentrations, and are on average
fainter, bluer, and less massive than early-type galaxies.

Here, we investigate whether the trends in optical fraction with 
$r_{\rm e}$/$R_{\rm 24}$ hold, even when some of the other properties, 
such as absolute magnitude $M_{\rm g}$, $g-r$ color, S\'ersic index $n$, 
and mass, are not allowed to vary significantly. To this effect,
we split our sample into 2 sub-groups according to the median values of $M_{\rm g}$ 
(Figure \ref{split}a), $g-r$ color (Figure \ref{split}b), S\'ersic index $n$ (Figure
\ref{split}c), and mass (Figure \ref{split}d). We then plot the optical bar
fraction as a function of $r_{\rm e}$/$R_{\rm 24}$ in each of the two subgroups, as
shown in Figures \ref{split}a to \ref{split}d. The solid points indicate bins
with more than 20 objects, whereas the open points represent bins with less
than 20 objects.

The optical bar fraction does not show any systematic variation between the
bright and faint subsamples at a given $r_{\rm e}$/$R_{\rm 24}$ (Figure
\ref{split}a). A similar result is seen with respect to mass (Figure
\ref{split}d), and color (Figure \ref{split}b).
This shows that even for samples with a narrow range in mass, color, or
luminosity, the trend of rising optical fraction with larger 
$r_{\rm e}$/$R_{\rm 24}$ (i.e. lower bulge-to-disk ratio) remains strong.
 
Figure \ref{split}c shows a significant difference in optical bar
fraction between the subsamples separated by S\'ersic index $n$. At a given
$r_{\rm e}$/$R_{\rm 24}$, the optical bar fraction is systematically higher,
typically by more than 20\%, for the subsample with the lower S\'ersic index
$n\leq1.48$ (Figure \ref{split}c). Since the S\'ersic index $n$ is low in pure
disk galaxies, this result supports our suggestion that disk-dominated galaxies
with a very low $B/D$ display a significantly higher optical bar fraction than
galaxies with prominent bulges. The potential implications of such a relation
on bar formation and disk stability are discussed in $\S$ \ref{discu}.

\section{Discussion}\label{discu}
\subsection{Implication for bulge formation models}
Hierarchical $\Lambda$ cold dark matter (CDM) models \citep[e.g.,][]{som99,col00,ste02} 
provide a good description of how DM behaves on large scales. 
By modeling the baryonic component and feedback processes, predictions
can be made regarding the disk, bulge, and bar components of galaxies.
In such models, gas with low angular momentum settles in the central parts of CDM halos to form 
small and dense protodisks \citep[e.g.,][]{whi78,don04}. Subsequent mergers of these
central stellar disks lead to classical spheroidal bulges with a de Vaucouleurs
$r^{1/4}$ profile \citep[e.g.][]{ste02,tay03}. It is also conceivable that the
bulge does not form in a single event, but is assembled by star forming clumps,
which originate in a proto-disk and coalesce in the center of the disk and form
a bulge \citep{nog99,imm04}. This possibility has gained support by the
observation of disk galaxies with prominent clumps at high redshift
\citep{elm04b}. The later accretion of high angular momentum gas around this
bulge invariably produces a spiral galaxy with a classical bulge and an
extended disk. In major mergers of spirals, violent relaxation destroys the
disks to produce an elliptical galaxy, while  minor mergers with mass ratios
above 1:4 typically spare the disk. 

While observations support many aspects of hierarchical $\Lambda$CDM models, 
the latter face several challenges. In particular, many cosmological
simulations with a merger history inclusive of major mergers fail to
produce galaxies without classical bulges
\citep[e.g.,][]{bur04,don06}, while high resolution simulations of
assembling disks with feedback from stellar energy
\citep[e.g.,][]{hel07} seem to reproduce a range of bulge-dominated to 
bulgeless disks. In this context, our study finds that in the range
$-18.5\leq M_g < -22.0$ mag and redshift $0.01\leq z<0.03$, $\sim$~ 20\% of the 
900 disk galaxies that are visually classified appear to be ``quasi-bulgeless'', 
without a classical bulge.  A similar fraction of 15\% for bulgeless galaxies  
was reported in the study of inclined disks by \cite{kau06}.

Another aspect of bulge formation not usually addressed by  hierarchical 
models is the formation of  disky bulges with high $v$/$\sigma$ (or `pseudobulges').
There are significant differences between classical bulges and disky bulges. 
While classical bulges form by gravitational collapse or hierarchical
merging, disky bulges are believed to form through gas inflows triggered  by bars or
any other non-axisymmetric feature in the gravitational potential
\citep{kor93,kor04,jog05,deb06}. Classical bulges are typically an order of
magnitude more massive than disky ones, therefore they are also brighter and
larger. Furthermore, studies of the stellar populations of classical bulges
indicate that their stars have been formed very quickly and long ago
\citep{pel99}. Our results show that more bars are present in late-type disks 
where typically  disky `pseudobulges' lie, a fact consistent with a 
bar-driven origin for `pseudobulges'.

\subsection{Implications for disk stability and bar formation scenarios}
One of the main results of our study is that the optical bar fraction
rises from $\sim45\%$ in early-type galaxies to a significantly
higher value ($\ge70\%$ ) in late-type galaxies, which appear
quasi-bulgeless, and seem to have a low bulge to disk ($B/D$) ratio,
as measured by $r_{\rm e}$/$R_{\rm 24}$, and confirmed by visual inspection.
The optical bar fraction shows a similar but shallower trend with
mass, rising in low mass galaxies (Figure \ref{massp}). Our conclusion
is also supported by \cite{ode96}, who finds that the frequency of bars
roughly doubles from Sc to Sm galaxies, using the RC3 bar classifications
and Hubble types.

In this section, we discuss how our findings can be
interpreted in the context of bar formation scenarios. As one moves
from early/intermediate type galaxies (e.g., Sa to Sc) to late-type
(e.g., Sd, Sm) systems, several important properties change: the gas
mass fraction in the disk rises, the $B/D$ ratio falls, and the total
mass falls. In addition, it is also found that the dark matter (DM)
fraction rises in lower luminosity systems
\citep[e.g.][]{per96,kas06}, but the scatter in such relations is
large \citep{kas06}. How do these changes along the Hubble sequence
impact the susceptibility  of a disk to form bars and the subsequent
bar evolution?

The higher gas  mass fraction present in late-type disks makes the 
disk dynamically cold and lowers the Toomre $Q$ parameter
\citep{too64}, defined as
\begin{equation}
\rm Q = \frac{ \kappa \ \sigma } { \pi \ \rm G \ \Sigma_{\rm disk}}
\end{equation}
where $\Sigma_{\rm disk}$ is the disk mass surface density, $\sigma$ is
the gas velocity dispersion, and $\kappa$ is the epicyclic frequency.
A low $Q$ (e.g., $\le$~2 to 3) favors the onset of bar instabilities  
and allows strong amplification in the context of the swing amplifier 
\citep{bin87}.

It has been proposed that the swing amplifier \citep{jul66,too81,bin87} model
may be relevant for bar formation. In such a scenario, a bar forms via a
resonant cavity of swing amplifying spiral density waves that reflect off
the center and the corotation radius. One way to suppress bar formation in
this model is to introduce an inner Lindblad resonance (ILR), which absorbs
the spiral density waves, thereby killing the feedback loop. In fact,
\cite{sel01} find that a disk with a sharp central density is completely
stable to bar formation. In the context of the swing amplifier,
a late-type galaxy with a low $B/D$ ratio would have a shallow rotation curve,
and may not harbor an ILR, thus favoring bar formation.

The DM halo can have a large impact on the growth of a bar.
The early work of \cite{ost73} suggested that the presence
of a dynamically important unresponsive DM halo can
suppress the bar instability in a disk galaxy.
However, the evolution of a bar is a highly non-linear process,
which depends on the exchange of angular momentum with the outer disk and
the DM halo via resonances (e.g., Weinberg 1985; Athanassoula 2002,
2003; Debattista \& Sellwood 1998, 2000; Berentzen, Shlosman, \& Jogee 2006;
Berentzen, Shlosman, \& Martinez-Valpuesta 2007).
Work with live halos has showed that there is resonant transfer
of angular momentum between the bar, the DM halo, and the outer disk
(e.g., Debattista \& Sellwood 1998, 2000; Athanassoula 2002; 2003): the
angular momentum absorbed by the DM halo makes an existing bar grow
and slow down. In fact, in the simulations with live halos of \cite{ath02},
a strong bar grows even in disks whose DM halo mass within the optical
radius exceeds that of the disk mass. It should also be noted that
even if the dark matter only becomes important outside the bar radius,
it can still interact with outer resonances in the bar potential, causing
the bar to grow. Thus, it appears that the larger dark matter fraction
in late-type disks would favor the growth of a bar, if one already
exists. It still remains unclear, however, whether a massive DM halo
would promote the formation of a bar in the case of an unbarred disk.
Furthermore, the shape of the DM halo (triaxial or axisymmetric) also
has an important impact \citep[e.g.,][]{ber06a,ber07}. We also note that
cosmological simulations of galactic disks \citep[e.g.,][]{gov07,hel07}
show extensive bar-forming activity, but there has not been any specific
prediction of how the bar fraction would vary as a function of Hubble type.

It is also important to consider whether our results could be
explained in terms of the evolution and destruction of bars
rather than their initial formation. Most simulations
\citep{she04,ath05,mar06,deb06} indicate that
{\it present-day} bars are relatively robust against the type of
central mass concentrations (CMCs) and $B/D$ that exist in {\it present-day}
galaxies today or in the recent past (see also the discussion in
$\S$~\ref{impl}). Specifically, as outlined in \cite{ath05} and \cite{she04}, the
super-massive black holes (SMBHs), central, dense stellar clusters, gaseous
concentrations, and inner parts of bulges, which exist in present-day galaxies,
fail significantly to generate the required CMCs for bar destruction. This does
not exclude, however, the possibility that at very early epochs (e.g., $z>1.5$)
when disks were still assembling, the different prevailing physical conditions
(e.g., large CMCs and gas inflows) might destroy bars (e.g., Bournaud et
al. 2005, but see Debattista et al 2006; Heller, Shlosman, \& Athanassoula 2007).
Lenses, which are preferentially found in early-type disks \citep{kor79,kor04},
are sometimes interpreted as dissolving bars. Within this framework,
our results of a higher optical bar fraction of bars in
quasi-bulgeless, late-type galaxies may reflect the fact that bars in
early-type galaxies were destroyed more frequently during their earlier
assembly. We note that in our study, the bar maximum ellipticity is
on average higher in faint quasi-bulgeless galaxies than in galaxies
with bulges (Figure \ref{visp2}d).

\subsection {Implication for the evolution of bars over the last 8 Gyr in
bright galaxies}\label{impl}
The evolution of the optical bar fraction with redshift is a subject
of active study. Early small studies reported that the optical fraction
of bars shows a striking decline at $z>\sim0.5$ \citep{abr99},
and undergoes a dramatic order of magnitude decline from $\sim29\%$
to below $1\%$ \citep{vdb00}.

Subsequent studies \citep{jog04,elm04a,sht03,zhe05} ruled out a
dramatic order of magnitude decline, and reported a fairly constant
optical fraction of strong bars or prominent bars ($\sim23\%$ to 30\%)
over $z=0.2$ to 1.0 (look-back times of 3 to 8 Gyr). The results of
such studies allowed, within the error bars, for modest factors of $\sim2$
variation in the optical fraction of strong bars or of all bars.
For example, for bright ($M_{\rm V}<-19.3$) disks, \cite{jog04} find a
rest-frame optical fraction of strong ($e_{\rm bar} \ge$~0.4) bars of
$\sim$~36\%~$\pm$~6\% at $z\sim0.2-0.7$, and $\sim$~24\%~$\pm$~4\% at
$z\sim0.7-1.0$, allowing a range of 42\% to 20\% for the optical
fraction of strong bars, and yielding an average values of  $\sim$~30\%.
A similar result was found for the completeness cut of $M_{\rm V}<-20.6$ 
\citep{jog04}. However, not much weight was given to a possible factor 
of $\sim2$ variation due to the small number statistics and due to 
redshift-dependent systematic effects that may cause an artificial
loss of optical bars in the higher redshift bins. These include the
increasing obscuration by dust and star formation, with the average
star formation rate increasing by a factor of $\sim4$ over $z\sim0.2$
to 0.8 (Jogee et al. 2007), the degradation of the PSF ($0\farcs09$)
from 300 to 680 pc, and the surface brightness dimming by a factor of
5 from $z\sim0.2$ to 0.8.

Recent studies using Cosmos data \citep{sht07} with a significantly
larger sample of bright (massive ($M \ge 10^{10} M_{\odot}$) galaxies report
a moderate decline by a factor of 2 or 3 in the optical fraction of
strong bars (from $30\%-35\%$ at $z\sim0.2$ to $9\%-17\%$ at $z\sim$~0.8).
They also report a decline in the optical fraction of (strong+weak) bars
from $\sim60\%$ at $z\sim0$ to $\sim22\%-31\%$ at $z\sim0.8$. If this
decline is not caused by the afore mentioned redshift-dependent
systematic effects, it implies that the frequency of both strong and
weak bars is lower at earlier times.

We can compare our globally-averaged optical bar fraction at $z\sim0$
($\S$ \ref{glob}) to the results at intermediate redshifts, but it is
critical to note two things. Firstly, these studies are carried out
in the rest-frame {\it optical} band and therefore should be compared
to the {\it optical} bar fraction at $z\sim0$. Secondly, comparisons
should be made for galaxies of the same luminosity or/and mass range.

We first compare the SDSS results with the study by \cite{jog04} on
strong bars, where they find a rest-frame optical fraction of strong
($e_{\rm bar} \geq 0.4$) bars of $\sim$~36\%~$\pm$~6\% at $z\sim0.2-0.7$,
and $\sim$~24\%~$\pm$~4\% at  $z\sim0.7-1.0$.
If we restrict our sample to galaxies with $M_g\leq-19.3$ mag and only
consider bars that are strong (ellipticity $\geq0.4$) and large enough
(semi-major axis $\geq1.5$ kpc) to be characterized via ellipse-fitting
out to $z\sim0.8$, we get an optical $r$-band fraction for strong bars of
$\sim34\%$. (For $M_g\leq-20.6$, a value of 31\% is obtained, but number
statistics are low and based on only 54 galaxies). The value of 34\%
is only slightly higher, by a factor of 1.4, compared to the value of
$\sim$~24\%~$\pm$~4\% seen in the higher redshift bin ($z\sim0.7-1.0$)
of the \cite{jog04} study. Thus, we find that once
the loss of bars due to poor resolution is taken into account, the data
are consistent with the optical fraction of strongly barred galaxies
suffering at most a decline by a factor of $\sim1.4$ out to $z\sim1$.
In fact, as discussed in MJ07, if one assumes a further modest loss of
optical bars due to increasing obscuration, the data may even allow for
a rise in the total fraction  of strong bars out to $z\sim1$.

The study by \cite{sht07} focuses on massive and bright
galaxies, with masses in the range $1 \times 10^{10}$ to $1 \times
10^{11}$ $M_{\odot}$, and $M_{\rm V}$ in the range $-21.2$ to $-23.7$ mag.
The SDSS sample of 2000 disk galaxies over $z=0.01$ to 0.03
has very few such bright galaxies (Figures \ref{magd} and
\ref{barf}c), and thus a comparison over the same luminosity range is
not possible. However, we can compare the optical bar fraction over the
mass range $1 \times 10^{10}$ to $3 \times 10^{10}$ $M_{\odot}$, where
the SDSS and \cite{sht07} data overlap.
The SDSS-based optical bar fraction over this mass range is $\sim47\%$
over $z\sim0.01$ to 0.03. This is similar to the value of $\sim60\%$
in the first redshift bin ($z\sim0.17$ to 0.37) of the \cite{sht07} study.
If we consider only bars that are large enough (semi-major axis $>1.5$ kpc) to
be reliably characterized via ellipse-fitting out to $z\sim0.8$, the
SDSS-based optical bar fraction falls from $\sim47\%$ to $\sim39\%$.
For comparison, the optical bar fraction is  $\sim25\%$ in the last bin
($z\sim0.60$ to 0.84). Thus, once the loss of bars due to poor resolution 
is taken into account, the observed value of $\sim25\%$ is consistent
with the optical bar fraction declining by at most a factor of (39\%/25\%) 
or $\sim1.6$ over this mass range.

\section{Summary and conclusions}\label{sum}
We have used the $r$-band images from the NYU-VAGC of a sample of 3692 galaxies
with $-18.5\leq M_g < -22.0$ mag and redshift $0.01\leq z<0.03$ to find and
characterize bars. While most studies of bars in the local Universe have been
based on relatively small samples that are dominated by bright early type (Sa
to Sc) galaxies with bulges, the present sample also includes many galaxies
that are disk-dominated and of late Hubble types. Furthermore, the sample is 
$\sim$~10 times larger and samples a larger volume than earlier local samples 
We used a color cut in the color-magnitude diagram to select $\sim2000$ disk
galaxies.  We cross-check that  S\'ersic cuts  would yield a similar sample.
We identify and characterize bars and disks using $r$-band images and
a method based on ellipse fits and quantitative criteria. The typical seeing
($1\farcs4$ or 290 to 840 pc over $0.01\leq z<0.03$) is adequate for resolving
large-scale bars, whose typical diameters are $\ge$ 2 kpc. Smaller nuclear bars
are not the focus of this study. After the standard procedure of excluding
highly inclined ($>60^{\circ}$) systems, we find the following results. 

\begin{figure}
\epsscale{1}
\plotone{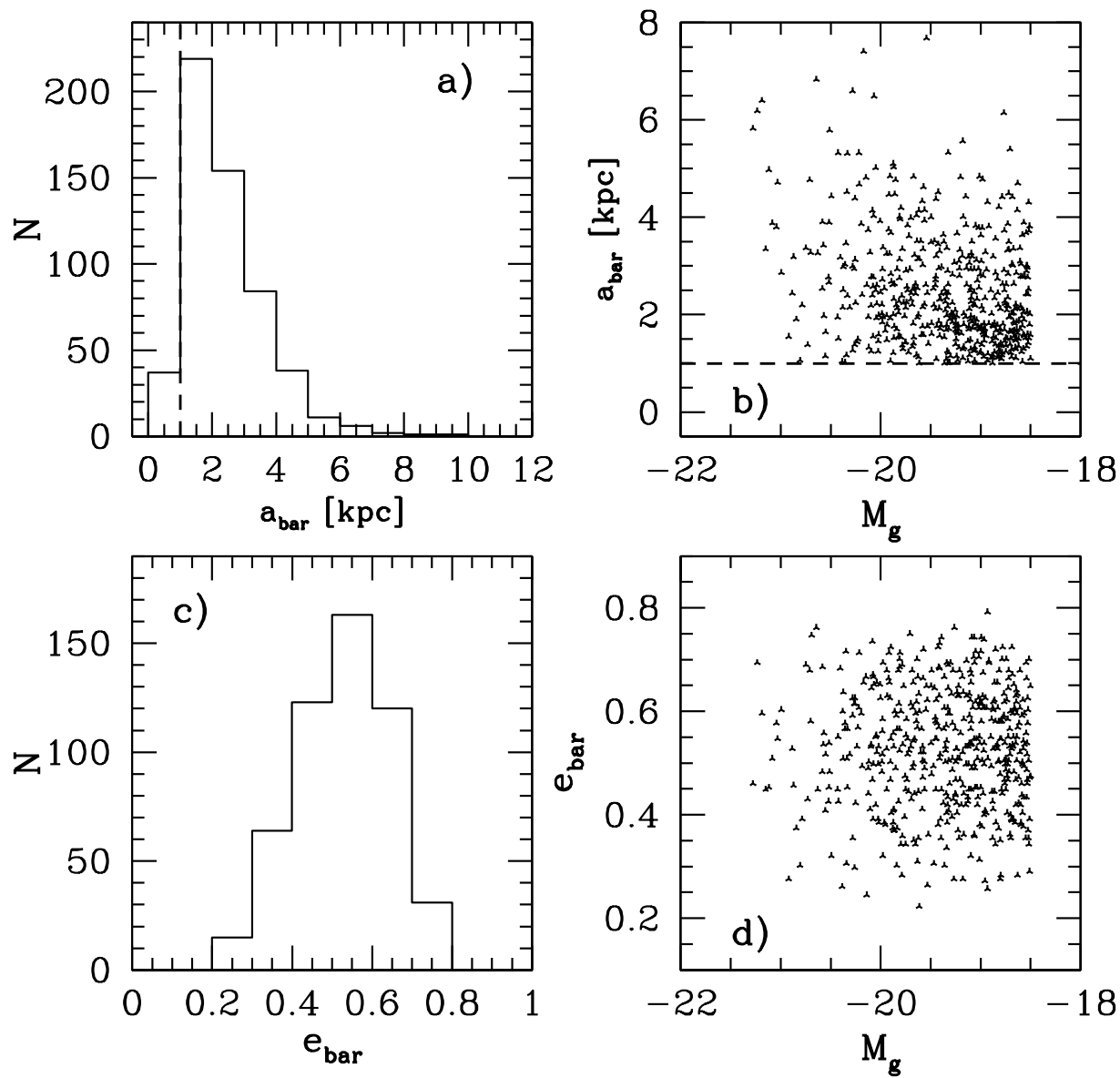}
\caption{{\bf a)} The distribution of bar semi-major axis lengths. We find very
few bars with sizes $>5$ kpc. {\bf b)} Plot showing absolute $g$-band magnitude
versus bar semi-major axis. {\bf c)} The distribution of bar ellipticities.
Most bars have ellipticities in the range 0.3 to 0.7. {\bf d)} Absolute $g$-band
magnitude versus bar ellipticity showing no obvious relation.\label{barp}}
\end{figure}

\begin{figure}
%\epsscale{2.3}
\plottwo{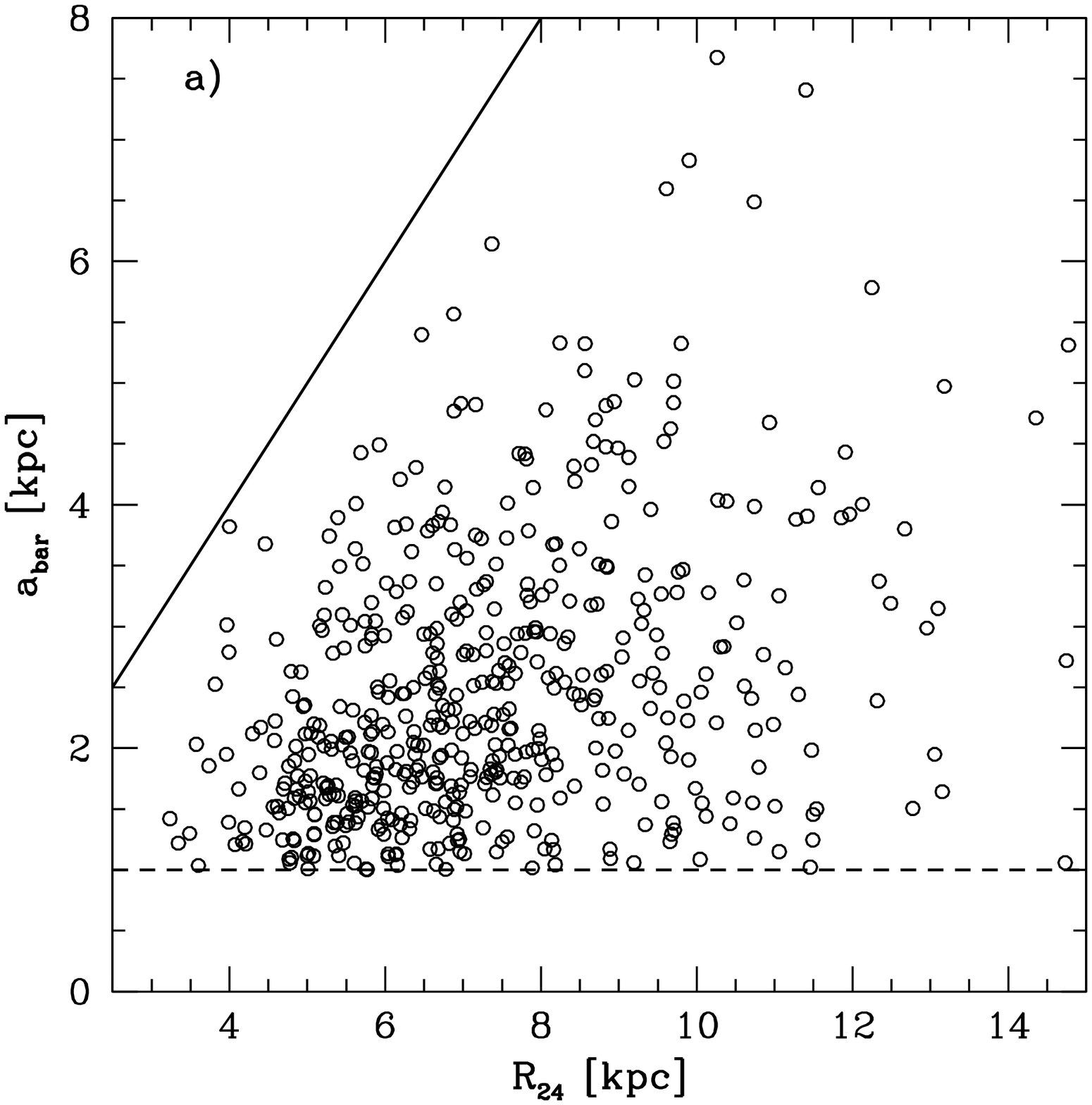}{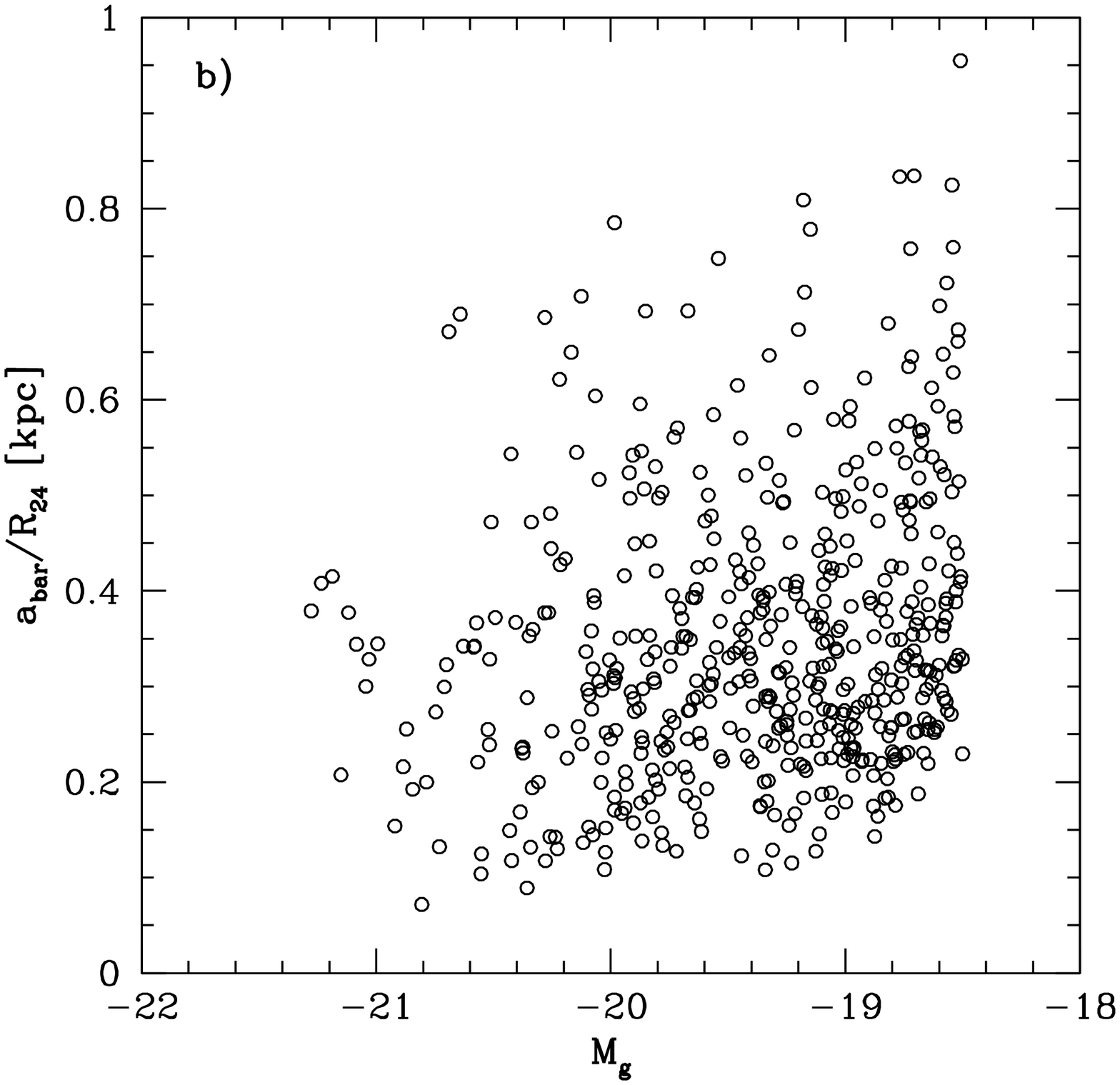}
\caption{{\bf a)} Plot of the bar semi-major axis versus $R_{24}$. The solid
line indicates $x=y$. There is no clear correlation between these two
parameters, but the bar typically ends inside $R_{24}$. {\bf b)} Absolute
$g$-band magnitude versus the ratio $a_{bar}/R_{24}$. Most galaxies have ratios
in the range 0.2 to 0.4 (median 0.32).\label{r24}}
\end{figure}

\begin{figure}
\epsscale{1}
\plotone{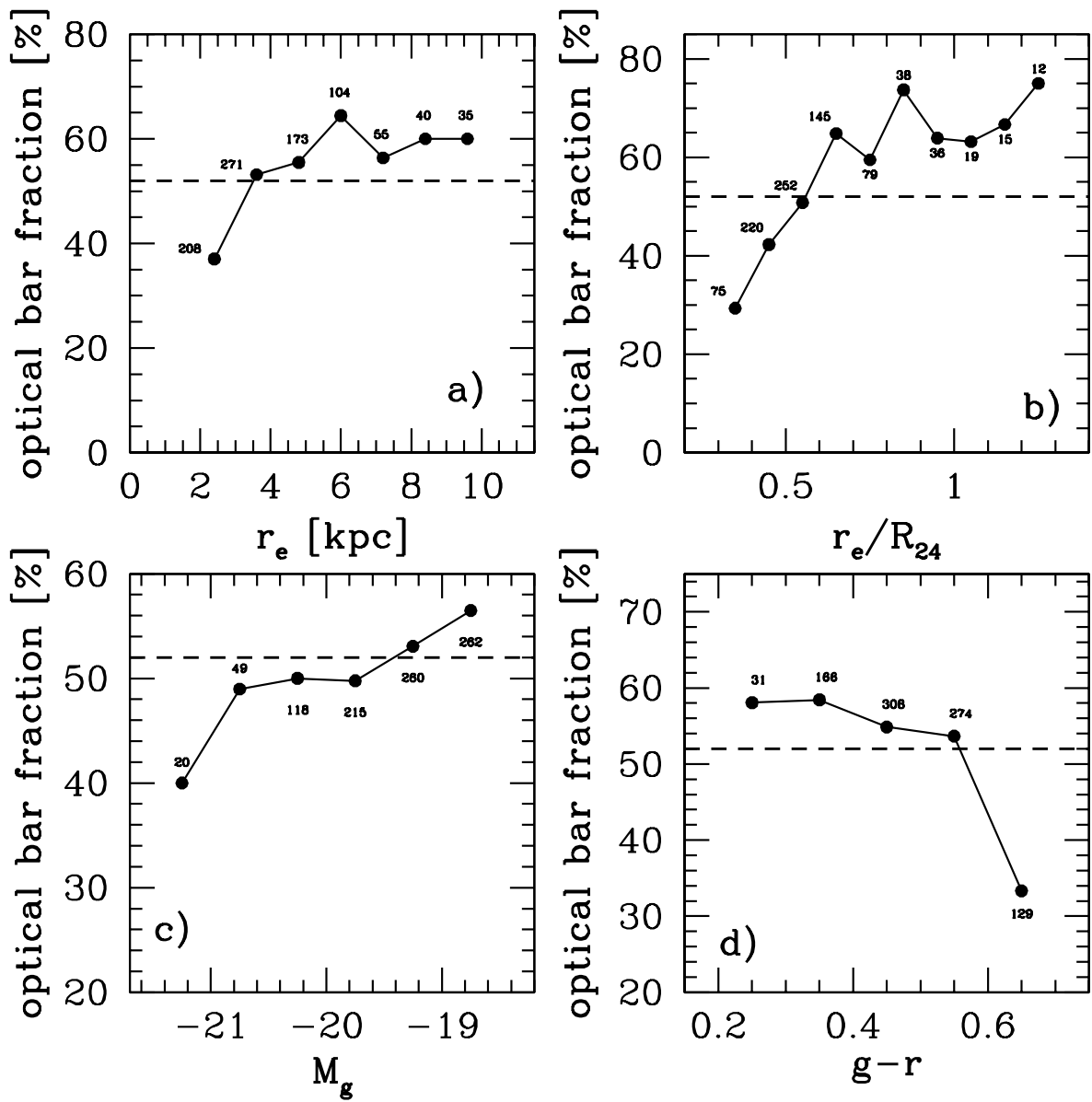}
\caption{The optical $r$-band bar fraction ($f_{\rm opt-r}$) as a function of
different galaxy properties. In all panels the numbers next to the points
indicate the number of galaxies in the corresponding bins. The dashed lines
indicate the total optical bar fraction ($52\%$). We only show bins with more
than 10 objects. {\bf a)} The optical bar fraction as a function of half light
radius $r_{\rm e}$: $f_{\rm opt-r}$ rises sharply, from $\sim40\%$ in galaxies
that have small $r_{\rm e}$ and visually appear bulge-dominated, to $\sim60\%$ for
galaxies that have large $r_{\rm e}$ ($\gtrsim4$ kpc) and appear disk-dominated.
{\bf b)} The optical bar fraction as a function of normalized
$r_{\rm e}$/$R_{\rm 24}$.  The smallest (or most compact) galaxies have a bar
fraction of $\sim30\%$, whereas the largest (most extended) galaxies reach a
value of $\sim70\%$. {\bf c)} The optical bar fraction as a function of
absolute $g$-band magnitude: $f_{\rm opt-r}$ is roughly constant at $M_g<-19.5$
mag (neglecting the brightest bin, which is very small) and increases towards
the fainter end of the magnitude range, reaching almost $60\%$ for the faintest
bin. {\bf d)} The optical bar fraction as a function of $g-r$ color: Notice the
sharp increase in bar fraction toward bluer colors.\label{barf}}
\end{figure}

\begin{figure}
%\epsscale{2.3}
\plottwo{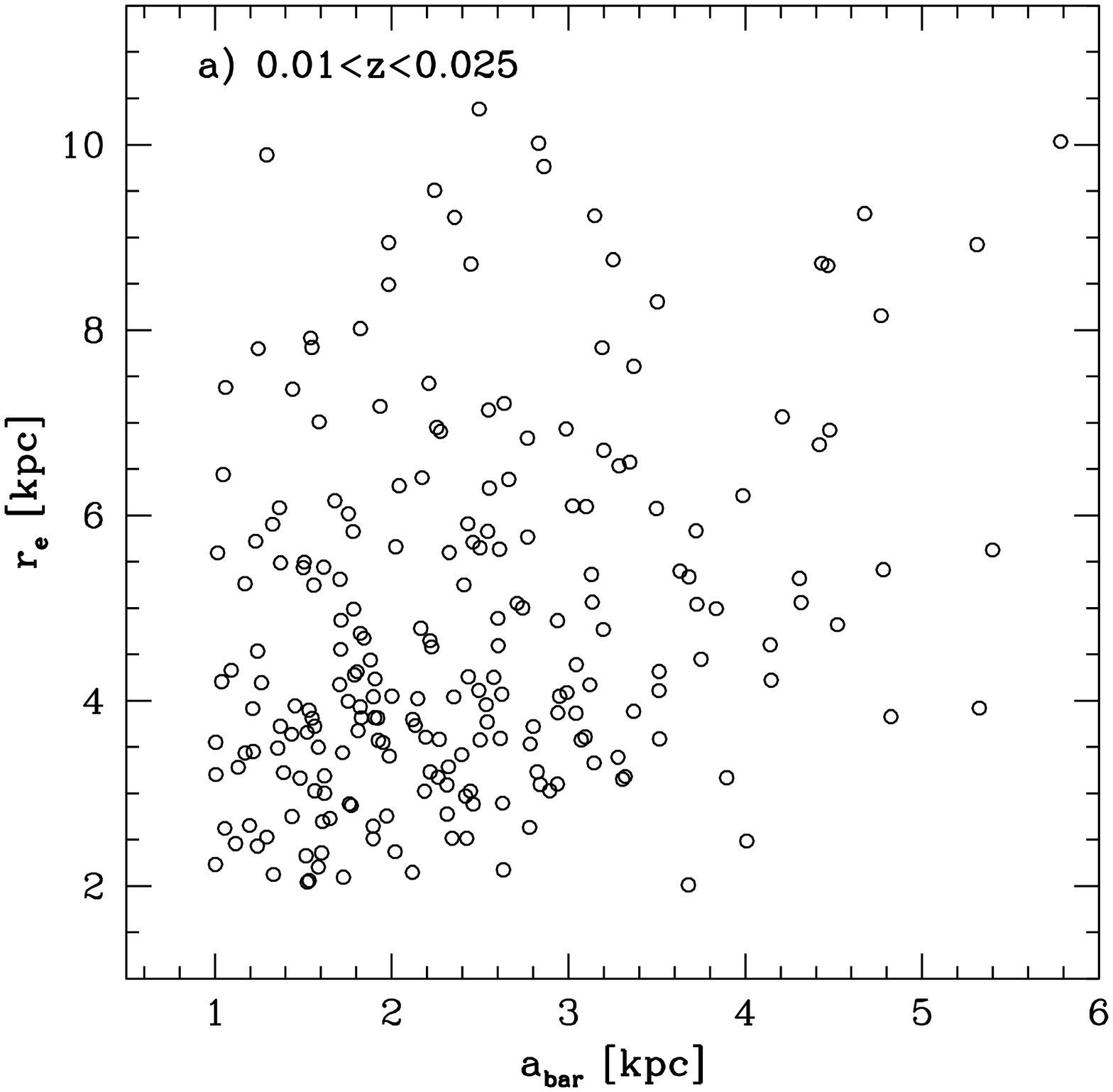}{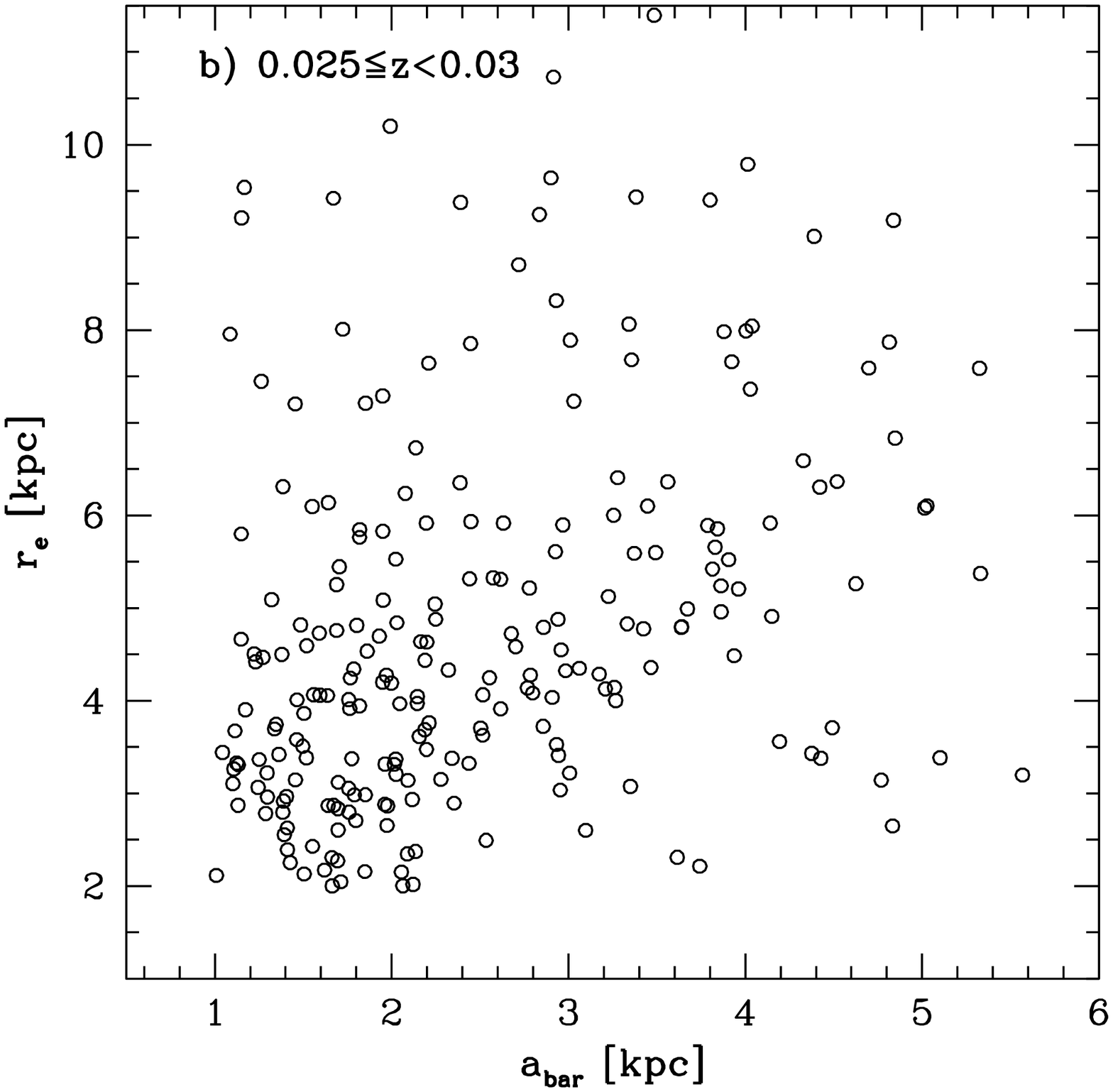}
\caption{Two plots showing the absolute bar size ($a_{bar}$) versus $r_{\rm e}$
for two redshift bins (panel a: $0.01<z<0.025$, panel b: $0.025\leq z <0.03$).
The two distributions are very similar and indicate that the bars in galaxies
with small $r_{\rm e}$ cover the whole range of bar sizes.\label{bsize}}
\end{figure}

\begin{figure}
\epsscale{1}
\plottwo{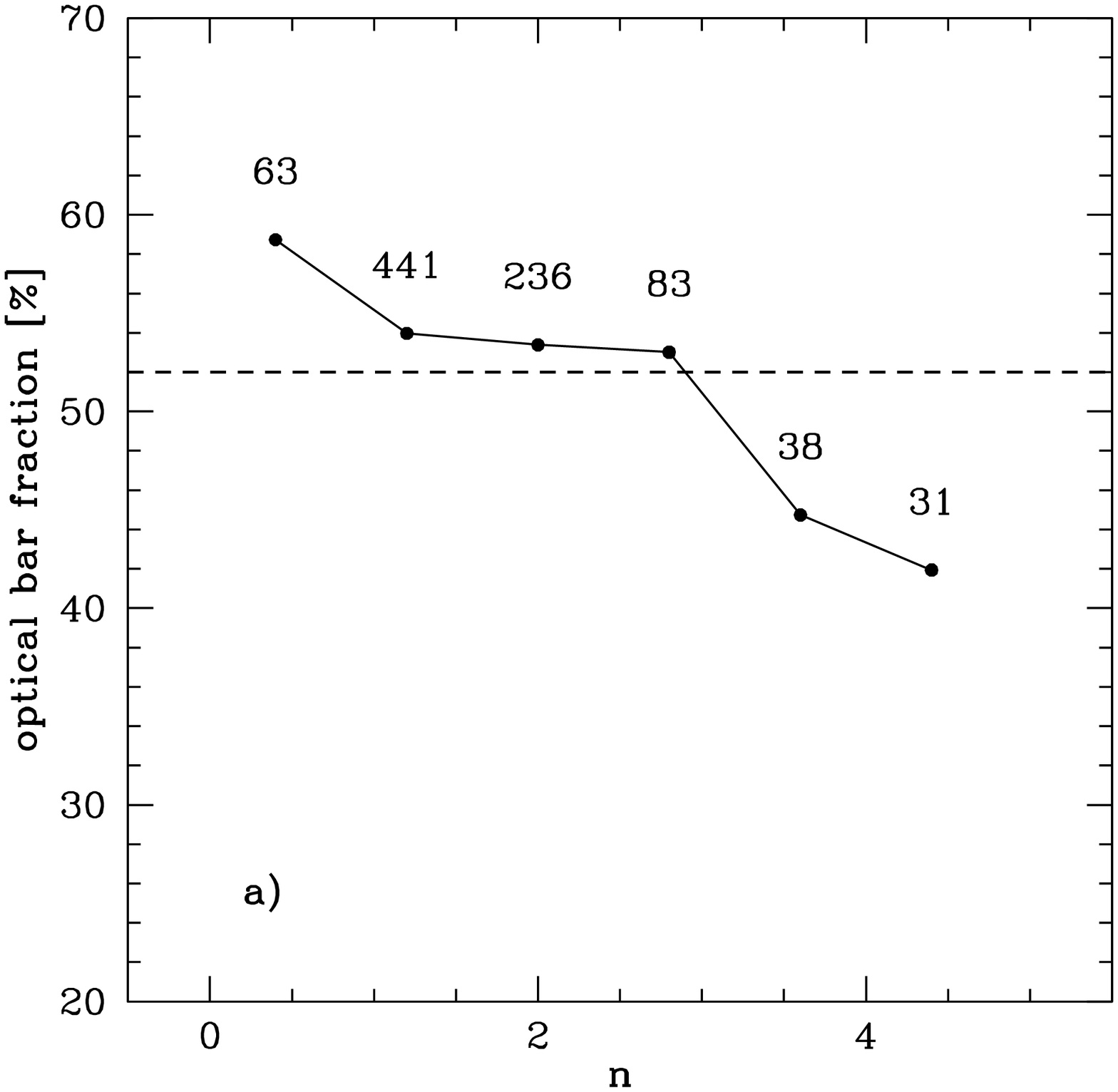}{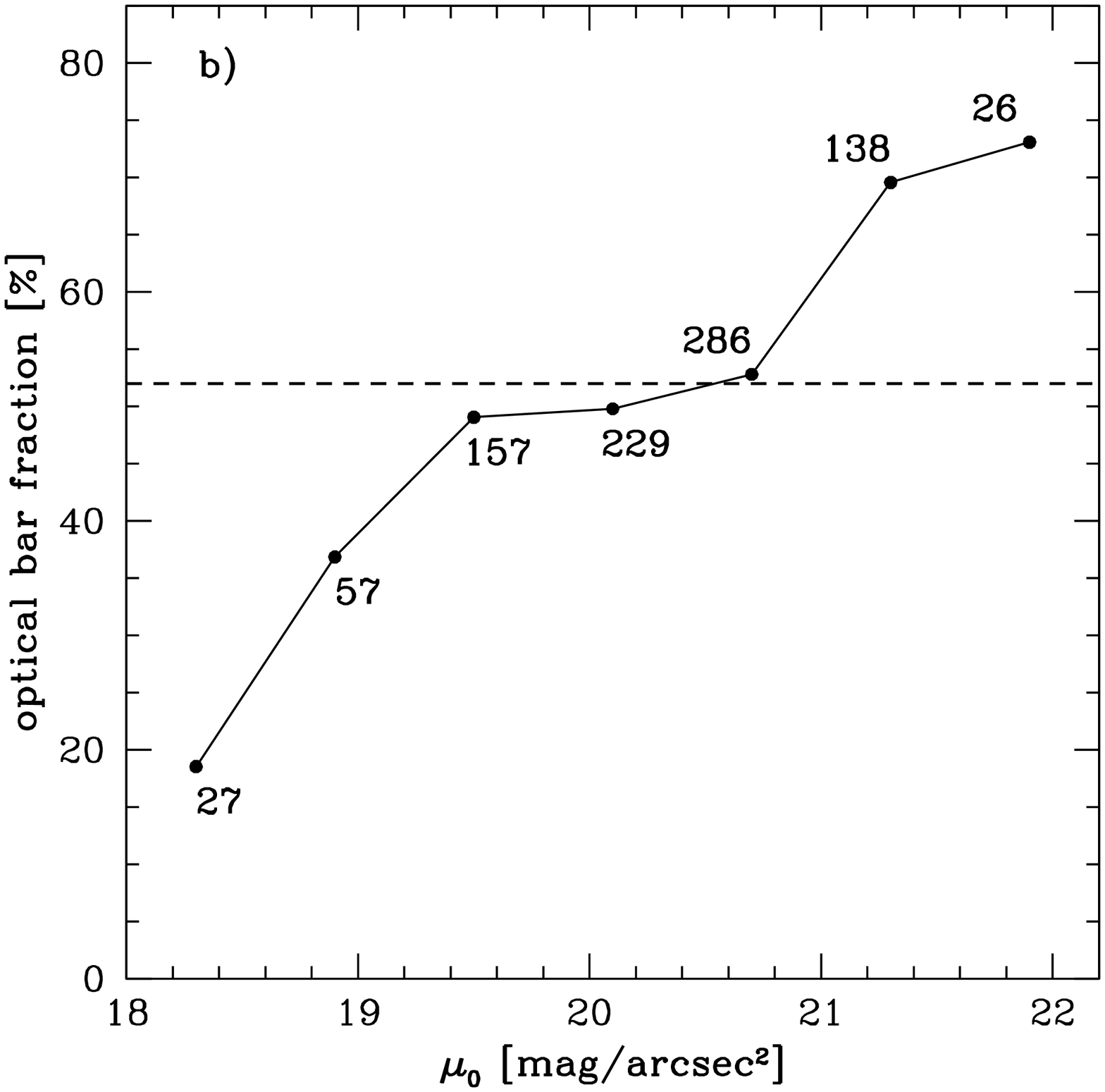}
\caption{
{\bf a)} The optical bar fraction as a function of $n$: the low $f_{\rm opt-r}$
for $n>$~2.5 are likely due to blue spheroids contaminating the color-selected
sample of disk galaxies. The rise in $f_{\rm opt-r}$ at $n\le$~1.5 is
consistent with a larger bar fraction in disk-dominated systems. {\bf b)} The
optical bar fraction as a function of $\mu_0$. The numbers in the plot and the
dashed line have the same meaning as in Figure \ref{barf}. The increase of the
bar fraction is not as steep as in the plot of $r_{\rm e}$ and
$r_{\rm e}$/$R_{\rm 24}$, but changes more continuously.\label{normfcsb}}
\end{figure}

\begin{figure}
\epsscale{1}
\plotone{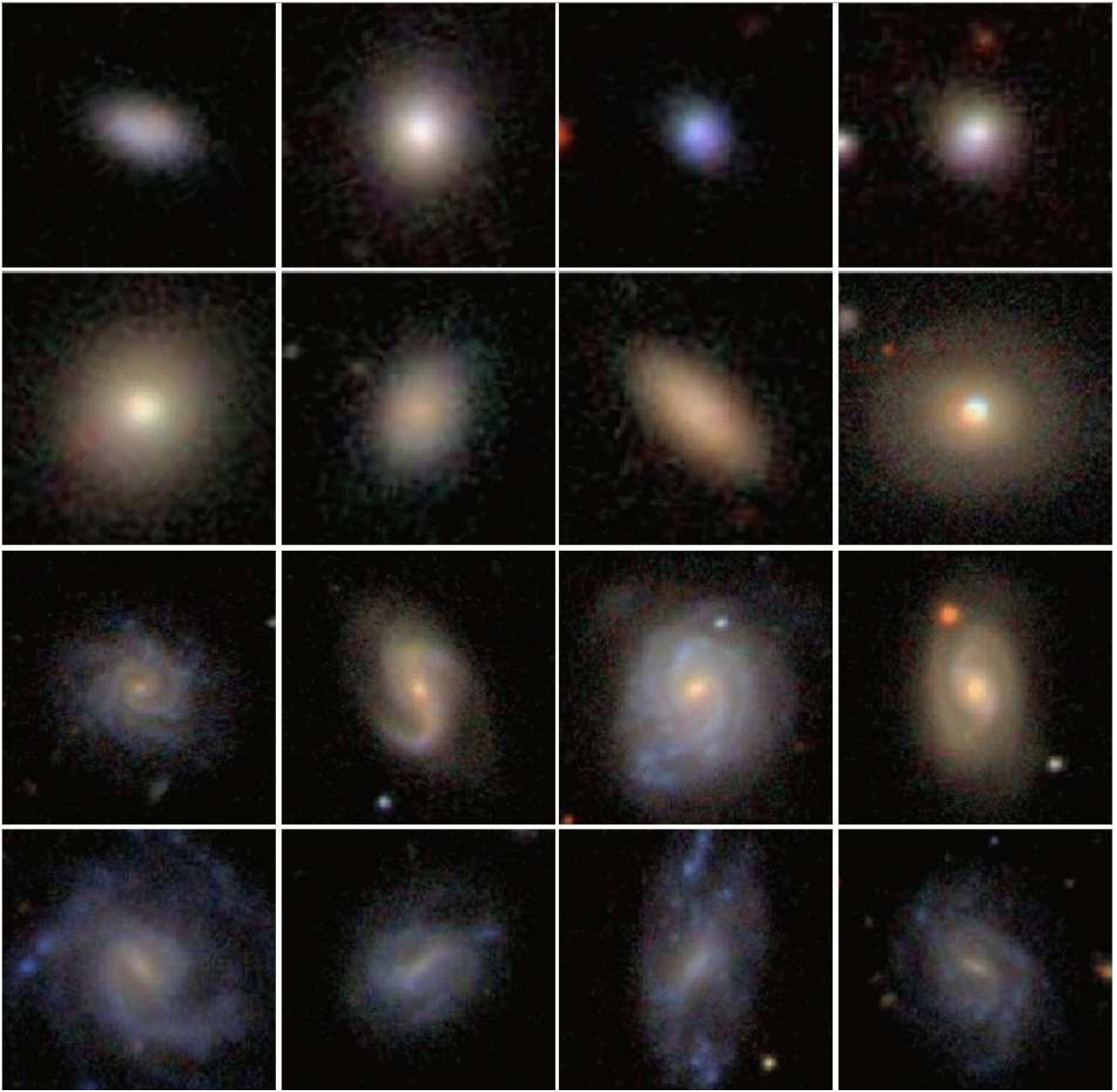}
\caption{Three band ($g$, $r$, $i$) color images from the SDSS archive for 16
objects from the test sample. The first row shows examples of objects, which
have been classified as pure spheroids (class 1), row two and three show
examples of objects with disk+bulge (class 2), and row four are pure disks
(class 3). The first seven images (from top and left to right) have a size of
$25\arcsec\times25\arcsec$ and the remaining nine have
$50\arcsec\times50\arcsec$.\label{ima}}
\end{figure}

\begin{figure}
\epsscale{1}
\plotone{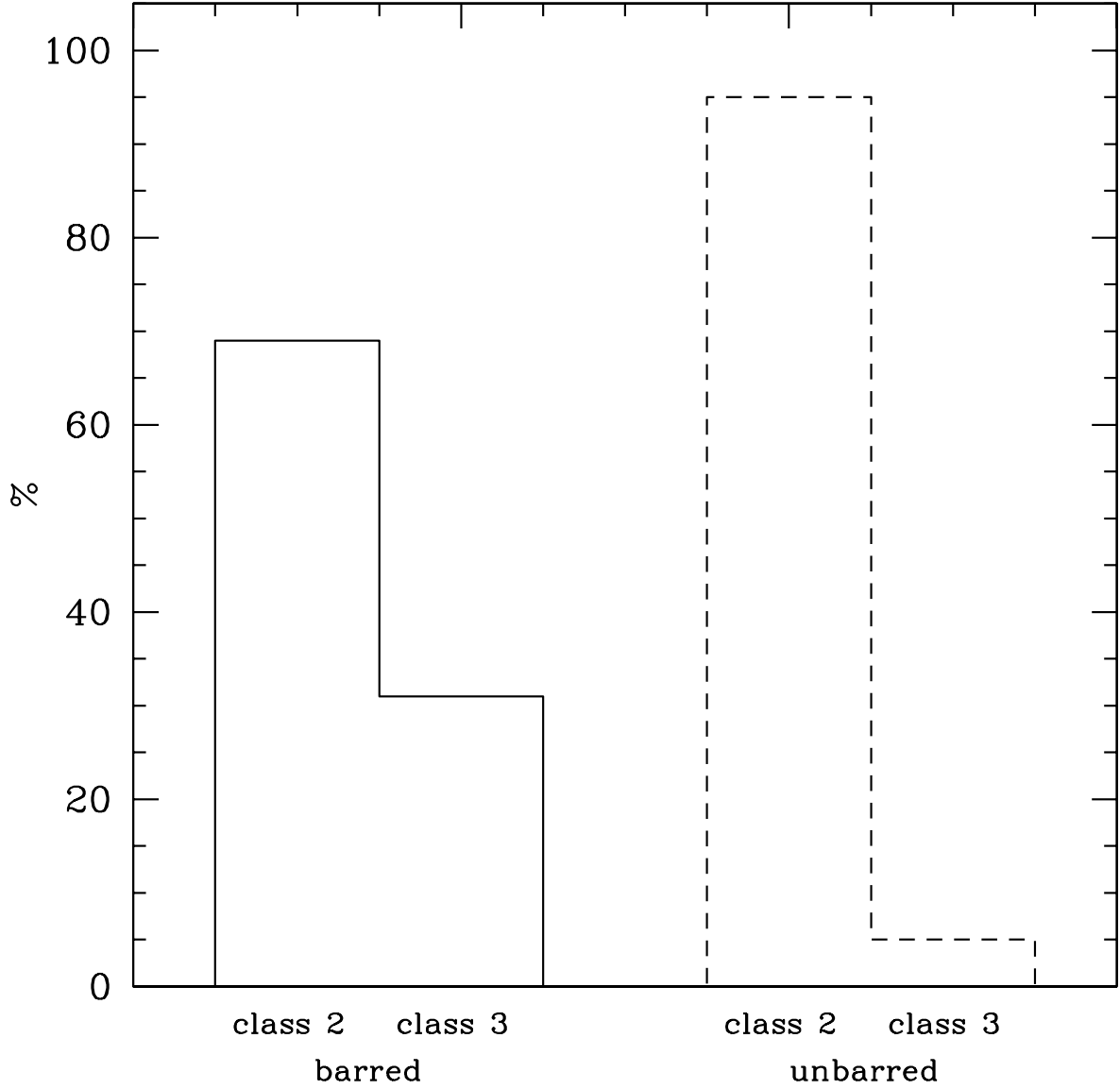}
\caption{The histograms show the percentage of class 3 (bulgeless) and
class 2 (bulge+disk) galaxies among barred galaxies (solid histogram) and 
unbarred galaxies (dashed histogram). The fraction of bulgeless galaxies in
much higher (31\% vs 5\%) in barred than unbarred systems.\label{visp}}
\end{figure}

\begin{figure}
\epsscale{1}
\plotone{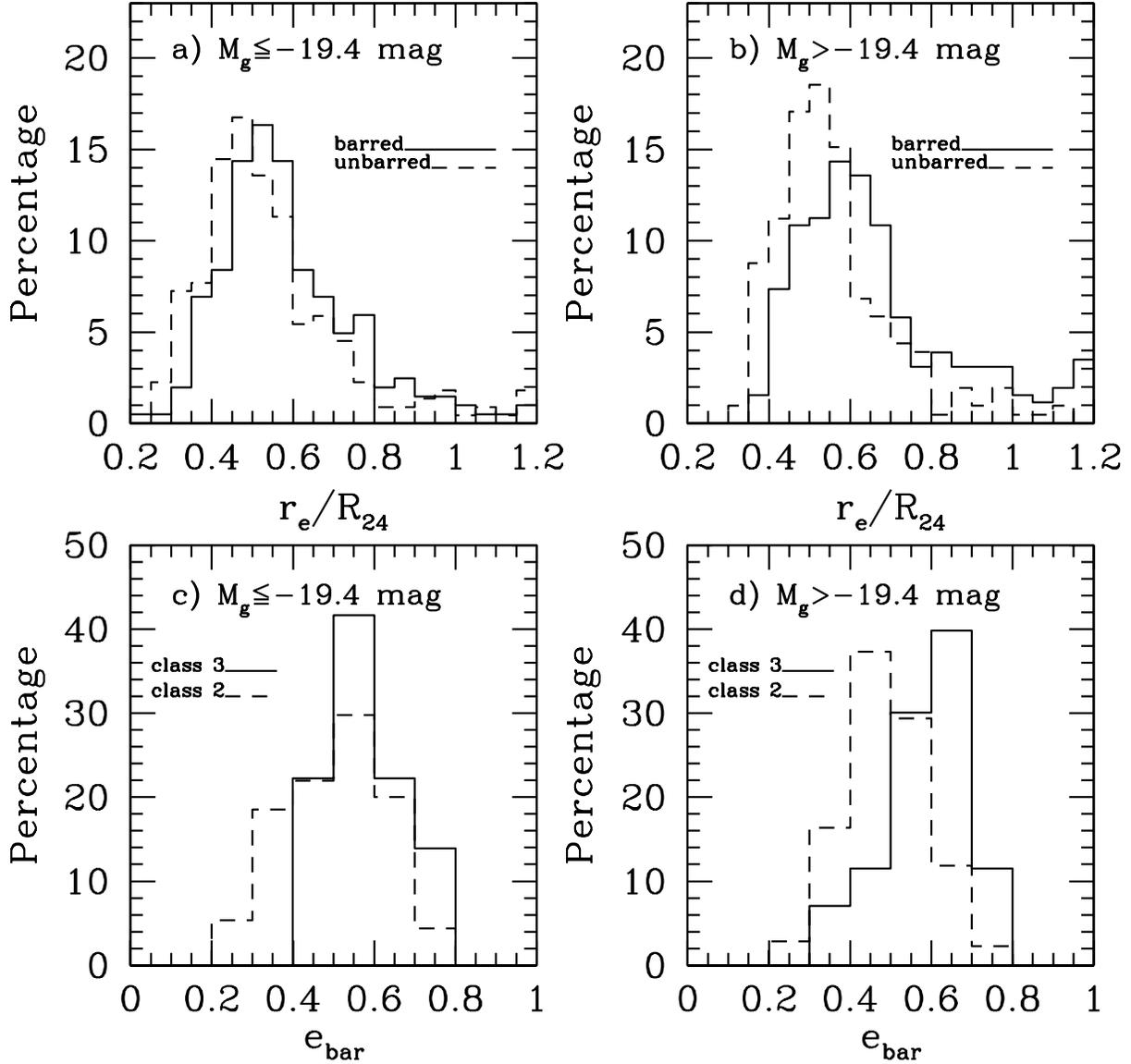}
\caption{{\bf a) and b)} The distribution of $r_{\rm e}$/$R_{\rm 24}$ among barred
galaxies (solid histogram) and unbarred galaxies (dashed histogram) for the
brighter galaxies (a) and the fainter galaxies (b). The fraction of galaxies
with large $r_{\rm e}$/$R_{\rm 24}$ ratios is higher in barred than unbarred
systems, particularly among the fainter galaxies. {\bf c) and d)} The
distribution of $e_{\rm bar}$ among galaxies in class 3 (Bulgeless; solid
histogram) and galaxies in class 2 (Bulge+Disk; dashed histogram) for the
brighter galaxies (c) and the fainter galaxies (d). The $e_{\rm bar}$
distributions in panel (d) indicate that bar ellipticities or strengths are on
average higher in faint  disk-dominated galaxies than in bulge-dominated
galaxies.\label{visp2}}
\end{figure}

\begin{figure}
\epsscale{1}
\plotone{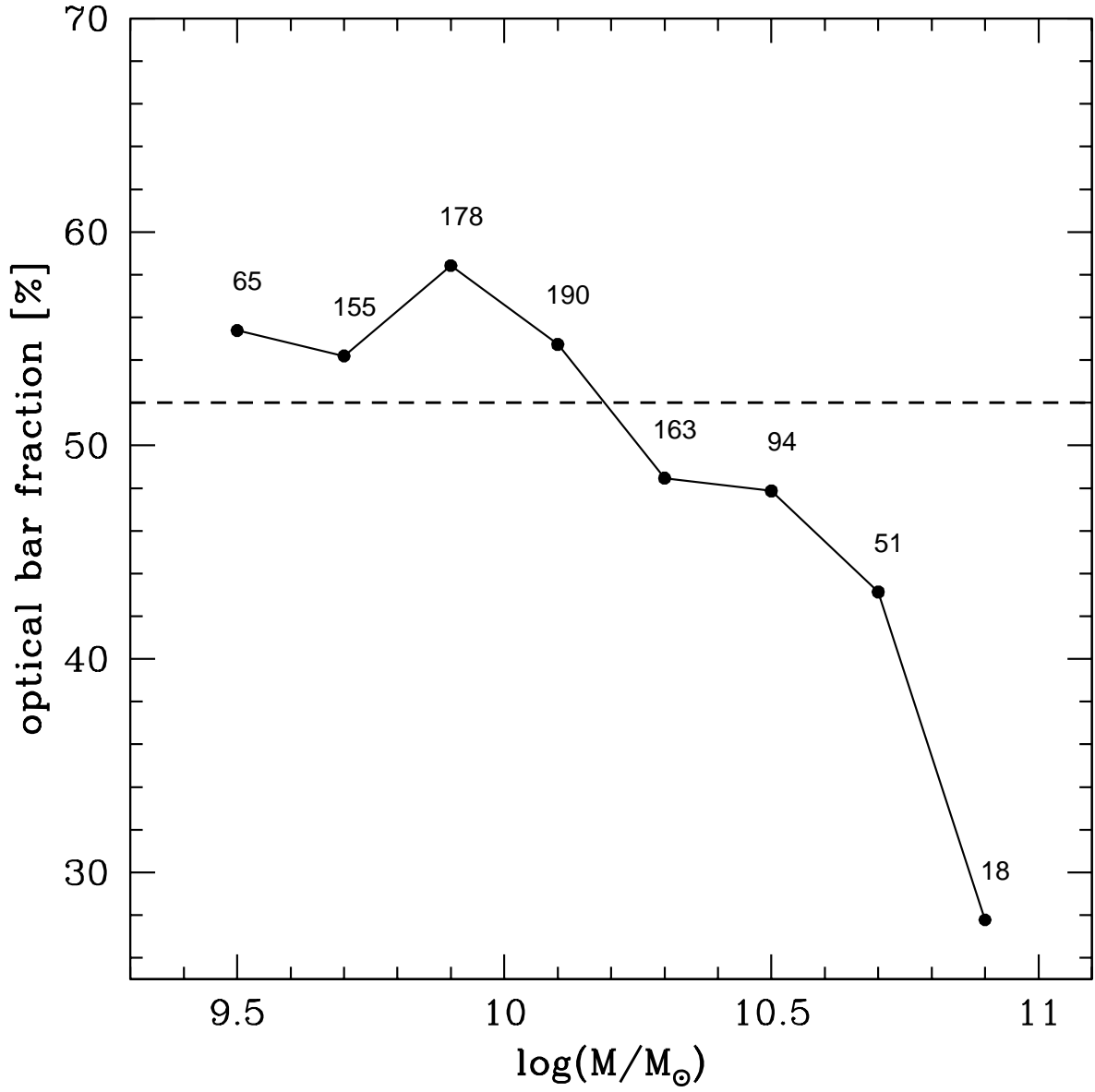}
\caption{The optical $r$-band bar fraction as a function of galaxy mass. The
masses have been determined using the $g-r$ color and the prescription of
\cite{bel03}.\label{massp}}
\end{figure}

\begin{figure}
\epsscale{1}
\plotone{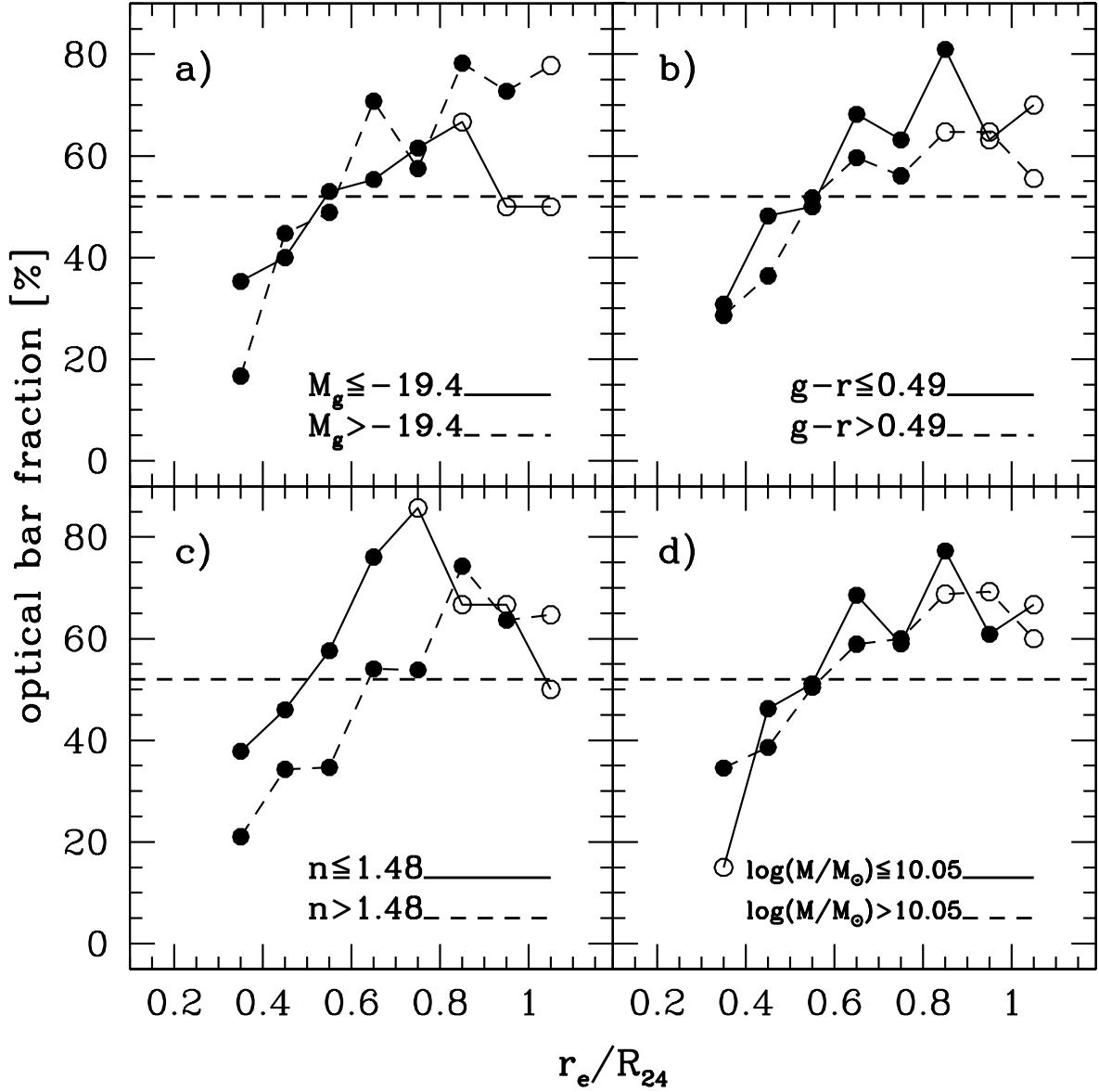}
\caption{The optical bar fraction as a function of $r_{\rm e}$/$R_{\rm 24}$ for
different subsamples defined according to the median values of basic galaxy
properties. The split locations are indicated in the four panels. The solid
points denote bins with more than 20 objects, whereas the open points represent
bins with less than 20 objects. The dashed lines indicate the total optical bar
fraction ($52\%$). The sample has been split based on {\bf (a)} absolute
$g$-band magnitude; {\bf (b)} $g-r$ color; {\bf (c)} S\'ersic index $n$; {\bf
(d)} galaxy mass.\label{split}}
\end{figure}

\begin{enumerate}
\item 
The average optical $r$-band bar fraction ($f_{\rm opt-r}$) in our sample,
which primarily consists of late-type disk-dominated galaxies, is $\sim48\%-52\%$.
The bars have diameters $d$ of 2 to 24 kpc, with most ($\sim72\%$) having
$d\sim$ 2 to 6 kpc (Figure \ref{barp}a). The bar length is typically much
smaller than $R_{24}$ (Figure \ref{r24}a) and most galaxies have a
$a_{bar}/R_{24}$ in the range 0.2 to 0.4 (Figure \ref{r24}b).

\item 
When galaxies are separated according to normalized $r_{\rm e}$/$R_{\rm 24}$,
which is  a measure of  the bulge-to-disk ($B/D$) ratio, a remarkable result is
seen: the optical $r$-band fraction rises sharply, from $\sim$~40\% in galaxies
that have small $r_{\rm e}$/$R_{\rm 24}$ and visually appear bulge-dominated, to
$\sim$~70\% for galaxies that have large $r_{\rm e}$/$R_{\rm 24}$.
Visual classification of $\sim80\%$ of our sample (with
$i<60^{\circ}$) confirms our result that
{\it late-type disk-dominated galaxies with
no bulge or a very low $B/D$ display a significantly higher optical bar
fraction ($>70$\% vs 40\%) than galaxies with prominent bulges.}
It also shows that barred galaxies host a larger fraction (31\% vs 5\%) of quasi-bulgeless
disk-dominated galaxies than do unbarred galaxies. The bar ellipticities or
strengths are on average higher in faint disk-dominated galaxies than in
bulge-dominated galaxies (Figure \ref{visp2}d). 

\item 
Similar trends in the optical bar fraction  
are found using the central surface brightness and color.
Bluer galaxies have higher  bar fractions ($\sim58\%$ at $g-r=0.3$) than
the redder objects ($\sim32\%$ at $g-r=0.65$) (Figure \ref{barf}d). 
The optical $r$-band fraction also shows a slight rise for galaxies
with fainter luminosities (Figure \ref{barf}c) and  lower masses
(Figure \ref{massp}). This is expected from (2), given that late-type
galaxies are fainter, bluer, and less massive.

\item 
The significant rise in the optical bar fraction toward disk-dominated
galaxies is discussed in terms of their higher gas mass fraction,
higher dark matter fraction, and lower bulge-to-disk ratio.

\item 
While many hierarchical $\Lambda$CDM models of galaxy evolution models fail 
to produce galaxies without classical bulges, our study finds that
in the range  $-18.5\leq M_g < -22.0$  mag and redshift $0.01\leq z<0.03$, 
$\sim20\%$ of the 1144 moderately inclined disk galaxies appear to
be ``quasi-bulgeless'',  without a classical bulge. 

\item Our study of bars at $z\sim0$ in the optical $r$ band provides a
reference $z\sim0$ baseline for intermediate redshift $HST$ ACS surveys that
trace bars in {\it bright} disks in the rest-frame optical bands ($BVRI$) out
to $z\sim1$. By applying the same cutoffs in magnitude, bar ellipticity
($e_{\rm bar} \geq0.4$), and bar size ($a_{\rm bar} \geq1.5$ kpc),
which are applied in $z\sim0.2-1.0$ studies in order to trace strong
bars with adequate spatial resolution in bright disks, we obtain an
optical $r$-band fraction for strong bars of $34\%$.
This is comparable to the values of $\sim30\%$ at $z\sim0.2-1.0$,
$\sim$~36\%~$\pm$~6\% at $z\sim0.2-0.7$, and $\sim$~24\%~$\pm$~4\% at
$z\sim0.7-1.0$. Our result implies that the optical fraction of strong bars
in bright galaxies does not suffer any dramatic order of magnitude decline
out to $z\sim$~1.
\end{enumerate}

\acknowledgments
F.D.B. and S.J. acknowledge support from the National Aeronautics and Space
Administration (NASA) LTSA grant NAG5-13063 and from HST-GO-10395 and
HST-GO-10428. S.J. and I.M. acknowledge support from the NSF grant
AST-0607748. We thank Isaac Shlosman, Juntai Shen, Lia Athanassoula, 
Victor Debattista, Francoise Combes, Bruce Elmegreen, and Alfonso Aguerri 
for useful discussions. This research has made use of NASA's Astrophysics 
Data System Service.

\clearpage

\begin{deluxetable}{lrrrr}
\tablecolumns{5}
\tablewidth{0pc} 
\tablecaption{Basic properties of the subsamples, resulting from the
classification of the 1860 color selected galaxies}
\tablehead{& \colhead{Barred} & \colhead{Unbarred} & \colhead{Too Inclined}
& \colhead{No Class} \\
& & & \colhead{$i>60^{\circ}$}}
\startdata
Number & 553 & 591 & 648 & 68 \\
Percentage & $29.7\%$ & $31.8\%$ & $34.8\%$ & $3.6\%$ \\
$\langle M_g \rangle$ [mag] & $-19.23$ & $-19.27$ & $-19.12$ & $-19.50$ \\
$\langle g-r \rangle$ [mag] & 0.48 & 0.49 & 0.51 & 0.49 \\
$\langle z \rangle$ & 0.025 & 0.026 & 0.025 & 0.025 \\
$\langle r_{eff} \rangle$ [kpc] & 4.13 & 3.12 & 4.63 & 4.62 \\
$\langle n \rangle$ & 1.40 & 1.52 & 1.18 & 1.43
\enddata
\label{basic}
\end{deluxetable} 

\clearpage

\clearpage
\thispagestyle{empty}
\setlength{\voffset}{-12mm}

\clearpage
\setlength{\voffset}{0mm}

\clearpage

\end{document}